\documentclass[aps,twocolumn,superscriptaddress]{revtex4-1}
\usepackage{amsmath}
\usepackage{amssymb}
\usepackage{amsfonts}
\usepackage[overload]{empheq}
\usepackage{graphicx} %package to manage images
\graphicspath{ {./Figs/} }
\usepackage[utf8]{inputenc}
\usepackage[english]{babel}
\usepackage[colorinlistoftodos]{todonotes}
\usepackage{mathtools}
\usepackage{float}
\usepackage{array,multirow}
\usepackage{bm}
\usepackage{cases}
\usepackage[position=top]{subfig}
\usepackage{pdfpages}
\makeatletter
\AtBeginDocument{\let\LS@rot\@undefined}
\makeatother
\newcommand{\meff}{M_{\text{eff}}}
\newcommand{\ms}{M_{\text{s}}}

\newcommand{\xidl}{\xi_{\text{DL}}}

\newcommand{\NM}{\hat{m}}

\newcommand{\Rahe}{R_\text{AHE}}
\newcommand{\Rphe}{R_\text{PHE}}

\makeatletter
\def\@fnsymbol#1{\ensuremath{\ifcase#1\or \dagger\or *\or \ddagger\or
   \mathsection\or \mathparagraph\or \|\or **\or \dagger\dagger
   \or \ddagger\ddagger \else\@ctrerr\fi}}
\makeatother
\begin{document}
\title{Sagnac interferometry for high-sensitivity optical measurements of spin-orbit torque}
\author{Saba Karimeddiny}    
    \thanks{These authors contributed equally}
    \affiliation{Cornell University, Ithaca, NY 14850, USA}
\author{Thow  Min  Cham}
    \affiliation{Cornell University, Ithaca, NY 14850, USA}
\author{Daniel C. Ralph}
    \email[Correspondence email address: ]{dcr14@cornell.edu}
    %\email[Correspondence email address: ]{email@institution.com}% Your name
    \affiliation{Cornell University, Ithaca, NY 14850, USA}  
    \affiliation{Kavli Institute at Cornell, Ithaca, NY 14853, USA}
\author{Yunqiu Kelly Luo}
    \thanks{These authors contributed equally}
    \affiliation{Cornell University, Ithaca, NY 14850, USA}
    \affiliation{Kavli Institute at Cornell, Ithaca, NY 14853, USA}

%PNAS
% \significancestatement{Sagnac interferometry is demonstrated to provide approximately a factor of 100 improvement in sensitivity for optical measurements of the magnetic orientation of magnetic thin films, compared to conventional magneto-optical Kerr effect (MOKE) imaging.  We use Sagnac interferometry to test the most widely-used electrical technique for measuring current-induced torques in magnetic films with perpendicular magnetic anisotropy -- the harmonic Hall (HH) technique.  We identify a flaw in the harmonic Hall method that can lead it to significantly overestimate current-induced torques.  The optical and electrical measurements can be reconciled if the magnetic deflections produced by current-driven deflection do not generate the same Hall signals that are created when exactly the same magnetic deflections are produced by an applied magnetic field.}

% \authorcontributions{S.K. and Y.K.L. devised the experiment, built the Sagnac apparatus, and performed the measurements. T.M.C. fabricated the devices. S.K. performed the data analysis. S.K., D.C.R., and Y.K.L. wrote the manuscript. All authors discussed the results and the content of the manuscript.}
% \authordeclaration{The authors declare no competing interests.}
% \equalauthors{\textsuperscript{1} S.K. contributed equally to this work with Y.K.L.}
% \correspondingauthor{\textsuperscript{2}E-mail: dcr14@cornell.edu}

% \keywords{Spin-Orbit Torques $|$ Sagnac Interferometry $|$ The Planar Hall Effect} 

\begin{abstract}
We adapt Sagnac interferometry for magneto-optic Kerr effect measurements of spin-orbit-torque-induced magnetic tilting in thin-film magnetic samples. The high sensitivity of Sagnac interferometry permits for the first time optical quantification of spin-orbit torque from small-angle magnetic tilting of samples with perpendicular magnetic anisotropy (PMA). We find significant disagreement between Sagnac measurements and simultaneously-performed harmonic Hall (HH) measurements of spin-orbit torque on Pt/Co/MgO and Pd/Co/MgO samples with PMA. The Sagnac results for PMA samples are consistent with both HH and Sagnac measurements for the in-plane geometry, so we conclude that the conventional analysis framework for PMA HH measurements is flawed. We suggest that the explanation for this discrepancy is that although magnetic-field induced magnetic tilting in PMA samples can produce a strong planar Hall effect, when tilting is instead generated by spin-orbit torque it produces negligible change in the planar Hall signal.  This very surprising result demonstrates an error in the most-popular method for measuring spin-orbit torques in PMA samples, and represents an unsolved puzzle in understanding the planar Hall effect in magnetic thin films.
\end{abstract}

%PNAS
% \begin{document}

\maketitle

%PNAS
% \thispagestyle{firststyle}
% \ifthenelse{\boolean{shortarticle}}{\ifthenelse{\boolean{singlecolumn}}{\abscontentformatted}{\abscontent}}{}
% \dropcap{S}
Spin-orbit torques (SOTs) \cite{Miron2011,Liu2012} are of interest for achieving high-efficiency manipulation of magnetization in magnetic memory technologies. SOTs are produced when a charge current is applied through a channel with strong spin-orbit coupling and generates a transverse spin current; this spin current can exert a spin-transfer torque on an adjacent ferromagnet (FM), allowing for low-power, electrical control of FM order. Memory cells with perpendicular magnetic anisotropy (PMA) are often preferred over their easy-plane counterparts because they may be fabricated at a higher density and are more resilient to stray magnetic fields or device heating. Accurate quantification of SOTs in PMA systems is therefore important for the development of future technologies.

Several techniques are commonly used to quantify SOTs in PMA heterostructures \cite{Pi2010,Liu2012,Liu2012PRL,Kim2012,Garello2013,Hayashi2014,Pai2016,Li2016}, yet these methods often exhibit significant quantitative discrepancies with one another. The most-commonly-used method for PMA samples, the harmonic Hall (HH) technique, measures the strength of spin-orbit torques by using second-harmonic Hall signals to detect current-induced magnetic deflections relative to the out-of-plane orientation \cite{Pi2010,Kim2012,Hayashi2014}. This method is attractive for its simplicity and has been employed in hundreds of published papers, but it sometimes produces discrepancies and even clearly-unphysical torque values when applied to samples with relatively strong planar Hall effects \cite{Woo2014,Torrejon2014,Lee2014,Lau_2017,Zhu2019,Zhu_2019}.
Members of our research group have recently suggested that calculating SOTs from PMA HH measurements by ignoring the expected signal from the planar Hall effect provides results for the SOTs in better agreement with HH measurements on samples with in-plane anisotropy \cite{Zhu_2019}.

Here, we test the influence of the planar Hall effect on HH measurements of PMA samples by comparing to high-sensitivity optical measurements of current-induced magnetic tilting performed simultaneously with HH measurements on the same samples. This work builds upon previous polar magneto-optic Kerr effect (p-MOKE) measurements that have been employed to quantify spin-orbit torques acting on samples with in-plane magnetic anisotropy \cite{Fan2014,Fan2016} and on PMA devices for which large magnetic fields were applied to force to magnetic orientation in-plane \cite{Montazeri2015,Chen2017}.  However p-MOKE signals are second-order in deflections from the perpendicular orientation, so measurements of spin-orbit-torque-induced small-angle deflections in PMA samples (directly analogous to the HH method) require a more sensitive method of optical detection.  For this we adapt Sagnac interferometry \cite{Xia2006, Fried2014}.

Our comparison between simultaneous optical and HH measurements demonstrates that the standard analysis framework for PMA HH measurements, which takes into account signals due to the planar Hall effect, is incorrect for samples in which the planar Hall effect is significant. These discrepancies can be explained  if magnetic tilting in PMA samples driven by spin-orbit torque does not generate a significant planar Hall signal, even though exactly the same tilting driven by applied magnetic field does.  This very surprising conclusion requires changing the framework for analyzing the most-popular technique for measuring spin-orbit torques in PMA samples and also, more fundamentally, it presents a puzzle reflecting that there is not yet a full understanding of the interactions among spin currents, charge currents, and ferromagnets. 

% Our comparison between simultaneous optical and HH measurements demonstrates that the standard analysis framework for HH measurements is incorrect for samples in which the planar Hall effect is significant, and in fact the standard analysis fails even to account self-consistently for the dependence of HH signals on the angle of in-plane magnetic field.  These discrepancies can be explained  if magnetic tilting in PMA samples driven by spin-orbit torque does not generate a significant planar Hall signal, even though exactly the same tilting driven by applied magnetic field does.  This very surprising realization requires changing the framework for analyzing the most-popular technique for measuring spin-orbit torques in PMA samples and also, more fundamentally, it presents a puzzle reflecting that there is not yet a full understanding of the  interactions among spin currents, charge currents, and ferromagnets. 

\section{Background}
\subsection{Spin-Orbit Torques}
For both the HH and Sagnac measurements, we model the current-induced magnetic deflections using the Landau-Lifshitz-Gilbert-Slonczewski (LLGS) equation within a macrospin approximation \cite{Ralph2008}
\begin{align}
\begin{split}
    \dot{\NM} &= \gamma \NM\times\frac{dF}{d\NM} \; + \; \alpha \NM\times \dot{\NM} \\ &\quad\quad + \tau^0_\text{DL}\NM\times\left(\hat{\sigma}\times\NM\right) 
     \;+ \;  \tau^0_\text{FL}\hat{\sigma}\times\NM
    \label{LLGS}
\end{split}
\end{align}
where $\NM$ is the normalized magnetic moment of the FM, $F$ is the free energy density of the FM, $\gamma=2\mu_B/\hbar$ is the gyromagnetic ratio with $\mu_B$ the Bohr magneton, $\alpha$ is the Gilbert damping parameter, and $\hat{\sigma}$ is the direction of spin polarization impinging on the FM. The last two terms are a result of the SOT and can be written as

% for a magnetization near the out-of-plane direction can be written as
\begin{align}
    \tau^0_\text{DL(FL)} =  \xi_\text{DL(FL)}\frac{\mu_B J_e}{e M_s t_\text{FM}}
    \label{taudef}
\end{align}
where $\xi_\text{DL(FL)}$ is the dimensionless SOT efficiency for the damping-like (field-like) torque, $J_e$ is the electric current density in the spin source layer applied in the $X$ direction, $M_s$ is the saturation magnetization of the FM, and $t_\text{FM}$ is the thickness of the FM layer.  The $\hat{X}$ and $\hat{Y}$ axes are defined as depicted in Fig.\ 1. In an amorphous-film system with high symmetry, we expect $\sigma \parallel \hat{Y}$ for a current that goes in the $X$-direction; we will use this assumption throughout.

For samples with the magnetic moment oriented out-of-plane, the effects of current-induced torques can alternatively be expressed in terms of current-driven effective magnetic fields.  The current-driven effective field in the $X$ direction will correspond to the damping-like torque, $\mu_0\Delta H_X = \mp \tau^0_\text{DL}/\gamma$ , where the $\mp$ corresponds to the magnetic orientations $m_Z = \pm 1$.  The current-induced effective field in the $Y$ direction will be the sum of the field-like spin-orbit-torque contribution and the {\O}rstead field, $\mu_0\Delta H_Y = \mu_0 H_\text{Oe} \pm \tau^0_\text{FL}/\gamma$.

\begin{figure*}[t!]
    \centering
    \includegraphics[width=\linewidth]{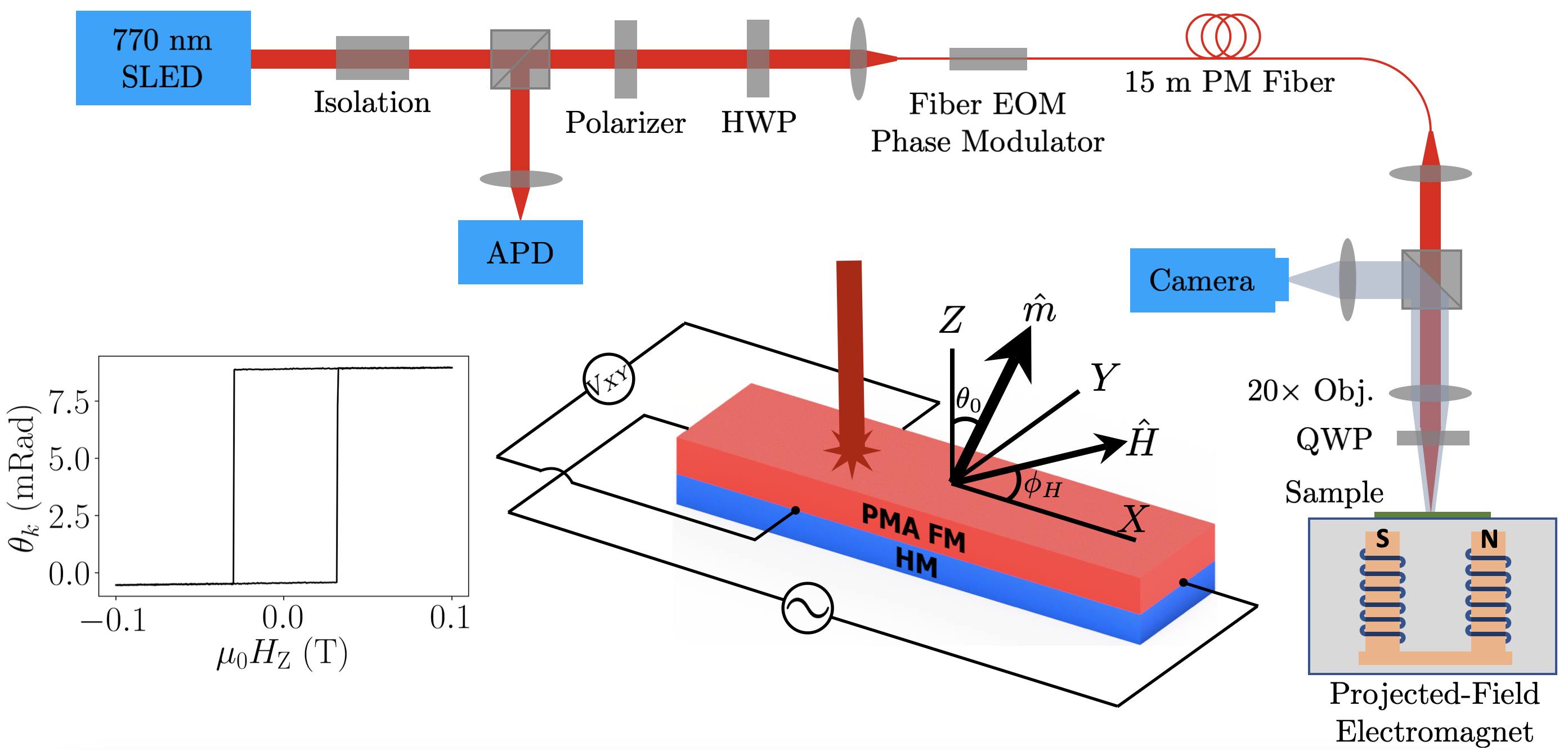}
    \caption{Schematic of the Sagnac interferometer. The left inset shows the Sagnac signal for magnetic-field-swept hysteresis of a Pt(4 nm)/Co(1.15 nm) device with $\mu_0 \meff \approx 0.42$ T; this is the same device for which we show data in Figs.\ \ref{PHECal} \& \ref{tilt}. The right inset depicts the device structure and coordinate definitions. In our measurements, $H$ is always applied in the $XY$\hspace{-0.08cm}-plane at $\phi_H=$ 0 or $\pi/2$. }
    \label{beamline}
\end{figure*}

\subsection{Harmonic Hall Measurement Technique}

We consider harmonic Hall (HH) measurements for a spin-source/ferromagnet bilayer in which the magnet has PMA and is initially saturated along the $\pm Z$-axis. A small external magnetic field, $H$, is applied in-plane at an angle $\phi_H = 0$ or $\pi/2$ relative to the $X$-axis using a projected-field magnet. In the absence of applied current, the equilibrium polar angle (measured from $Z$-axis) of the magnetization, $\theta_0$, can be written to good approximation as $\sin\theta_0 = H/\meff$ where the effective magnetization, $\mu_0\meff = 2K_\perp/\ms - \mu_0 \ms$, is the out-of-plane anisotropy minus the saturation magnetization; with this definition $\meff$ is a \textit{positive} quantity for a magnet with PMA.  A low-frequency (non-resonant) AC  voltage, $V(t) = \Delta I R_{XX} \sin\omega_e t$ [$\omega_e = 3137\;(2\pi) \text{s}^{-1}$ in our measurements], is applied to the device along the $X$-axis to generate deflections of the magnetic moment that can be characterized by current-induced effective fields $\Delta H_X$ and $\Delta H_Y$. The Hall voltage along the $Y$-axis is measured. 

For a system with a conducting magnet, the Hall resistance can depend on the magnetization orientation via both the anomalous Hall (AHE) and planar Hall effects (PHE),  $R_{XY} = R_{\text{PHE}}m_X m_Y + R_{\text{AHE}} m_Z$. Given the AC current in the $X$ direction, the Hall voltage will have a contribution at the drive frequency $\omega_e$ associated with the equilibrium magnetic orientation and a second-harmonic signal at $2\omega_e$ due to mixing between the AC current and the oscillations in $R_{XY}$ produced by the magnetic deflections.  For $\phi_H = 0$ or $\pi/2$, and within a small-angle approximation for $\theta_0$  \cite{Hayashi2014},  
%The Hall voltage is measured with a lock-in amplifier to discriminate the first- and second-harmonics, which are calculated (using a small-angle approximation for $\theta_0$) to be
\begin{align}
    V^\omega_{XY} = &\pm\Rahe\left(1 - \frac{H^2}{2M_\text{eff}^2}\right)\Delta I \label{transportEq1} \\
    V^{2\omega}_{XY} = &\left[\pm\Rahe\left(\Delta H_X\cos\phi_H + \Delta H_Y\sin\phi_H\right) \right. \nonumber\\
    & \left. -\Rphe\left(\Delta H_X\sin\phi_H + \Delta H_Y\cos\phi_H\right) \right]\nonumber \\
    & \times \frac{H}{2M_\text{eff}^2}\Delta I, \label{transportEq2}
\end{align}
where the $\pm$ accounts for magnetic saturation along the $\pm Z$-axis.
%These equations and their derivations can be found in ref.\ \cite{Hayashi2014} and an alternative derivation can be found in the supplemental information of this work \cite{Supplement}.  

%In a conventional HH measurement, the external magnetic field is swept along the current direction ($\phi_H = 0$) and perpendicular to the current direction ($\phi_H = \pi/2$).  
The current-induced effective fields $\Delta H_X$ and $\Delta H_Y$ acting on the out-of-plane magnetic moment can then be calculated as \cite{Hayashi2014} 
\begin{align} 
    \Delta H_{X} &= -2\frac{D_{0} \pm \epsilon D_{\pi/2}}{1-\epsilon^2} \label{TransportHX} \\
    \Delta H_{Y} &= -2\frac{D_{\pi/2} \pm \epsilon D_{0}}{1-\epsilon^2} \label{TransportHY}
\end{align}
where
\begin{align}
    D_{\phi_H} &= \frac{dV^{2\omega}_{XY}(\phi_H)}{dH}\left(\frac{d^2V^{\omega}_{XY}(0)}{dH^2}\right)^{-1}. \label{Bdef}
\end{align}
and $\epsilon = \Rphe/\Rahe$. (These results are consistent with ref.\ \cite{Hayashi2014} because our variable $R_\text{AHE}$ is equal to $\Delta R_A/2$ in ref.\ \cite{Hayashi2014} and hence our variable $\epsilon$ is equal to $2\xi$ in ref.\ \cite{Hayashi2014}.)  

\subsection{Sagnac MOKE Interferometry Technique}
 
In our experiments we remain below the maximum values of $\theta_0 < 0.25 \text{ Rad}$ and $\Delta\theta <$ 10 mRad.  Given a typical value of the Kerr rotation angle upon full reversal of a 1 nm PMA Co film ($2\kappa = \theta_k(\pi) - \theta_k(0) \sim$ 9 mRad, see Fig.\ 1) and that for small-angle-deflections from an out-of-plane configuration $\theta_k = \kappa m_Z$ so that the change in polar angle has a maximum value $|\Delta \theta_k| \approx \kappa \sin(\theta_0) \Delta \theta$, the oscillations in Kerr angle associated with the current-induced deflections are at most about 20 $\mu$Rad.
To achieve the sensitivity necessary to measure such small signals, we adapted a Sagnac interferometer design \cite{Xia2006,Xia2006_2,Gong2017} able to measure Kerr rotation with noise less than  $5\;\mu$Rad/$\sqrt{\text{Hz}}$.  
%This is sufficient to make a quantitative comparison with the electrical HH measurements of spin-orbit torques. 
The design of the Sagnac MOKE apparatus is described in Methods, and we compare the performance of conventional MOKE with our Sagnac apparatus in  Supplementary Information section VIII \cite{Supplement}.

 For measurements of current-induced torques with the Sagnac interferometer, we perform Sagnac and HH measurements simultaneously on the same samples to make sure that any effects of the LED illumination do not cause differences between the two techniques.  We therefore apply the same low-frequency AC voltage drive (at frequency $\omega_e$) as in the HH experiments and detect the time-varying signal MOKE signal from the interferometer  demodulated by a lock-in amplifier at both the driving frequency $\omega$ of the electro-optic phase modulator and (separately) at the lower-sideband frequency $\omega - \omega_e$ (see Supplementary Information section II for details \cite{Supplement}). 
 The signals at these frequencies measure the DC Kerr rotation ($\theta_k$) associated with the magnetic-field-induced equilibrium tilt angle ($\theta_0$) and the oscillations in the Kerr signal ($\Delta\theta_k$) associated with current-induced tilt ($\Delta \theta$), respectively. 
%Qualitatively, the $\theta_k$ signal will give us the same information as contained in the $V^\omega_{XY}$ signal from the HH experiment and the $\Delta\theta_k$ signal will give us the same information as contained $V^{2\omega}_{XY}$. Indeed, t
The expected Sagnac signals have the form
\begin{align}
    \theta_k &= \pm\kappa\left(1 - \frac{H^2}{2\meff^2}\right) \label{ACthetak}\\
    \Delta\theta_k &= \pm\kappa\left(\Delta H_X \cos\phi_H +\Delta H_Y \sin\phi_H\right) \frac{H}{\meff^2}\label{ACdeltathetak}.
\end{align}
 Here $\kappa$ is the constant of proportionality that relates the out-of-plane component of magnetization to the Kerr rotation, analogous to $\Rahe$ for the electrical measurement. There is no MOKE contribution that acts like the PHE in equation (\ref{transportEq2}) because Sagnac signal has negligible dependence on the in-plane components on the magnetic moment (see Supplementary Information section IV \cite{Supplement}).  Based on these equations, for a PMA sample the component of the current-induced effective fields are simply 
\begin{align}
    \Delta H_{X} &= -\frac{d\Delta \theta_k (\phi_H=0)}{dH}\left(\frac{d^2\theta_k}{dH^2}\right)^{-1} \label{HXSagnac} \\
        \Delta H_{Y} &= -\frac{d\Delta \theta_k (\phi_H=\pi/2)}{dH}\left(\frac{d^2\theta_k}{dH^2}\right)^{-1}. \label{HYSagnac}
\end{align}

\section{Experimental Results}
We will present measurements on two series of samples: Substrate/Ta(1.5)/Pt(4)/Co(0.85$-$1.3)/MgO(1.9)/Ta(2) and Substrate/Ta(1.5)/Pd(4)/Co(0.55$-$0.65)/MgO(1.9)/Ta(2) heterostructures where the numbers in parentheses are thicknesses in nanometers. Studying devices with different Co-layer thicknesses allows us to tune the strength of the out-of-plane magnetic anisotropy. The Hall-bar devices measured are 20 $\mu$m $\times$ 80 $\mu$m in size. 

\begin{figure}[h!]
    \centering
    \includegraphics[width=\linewidth]{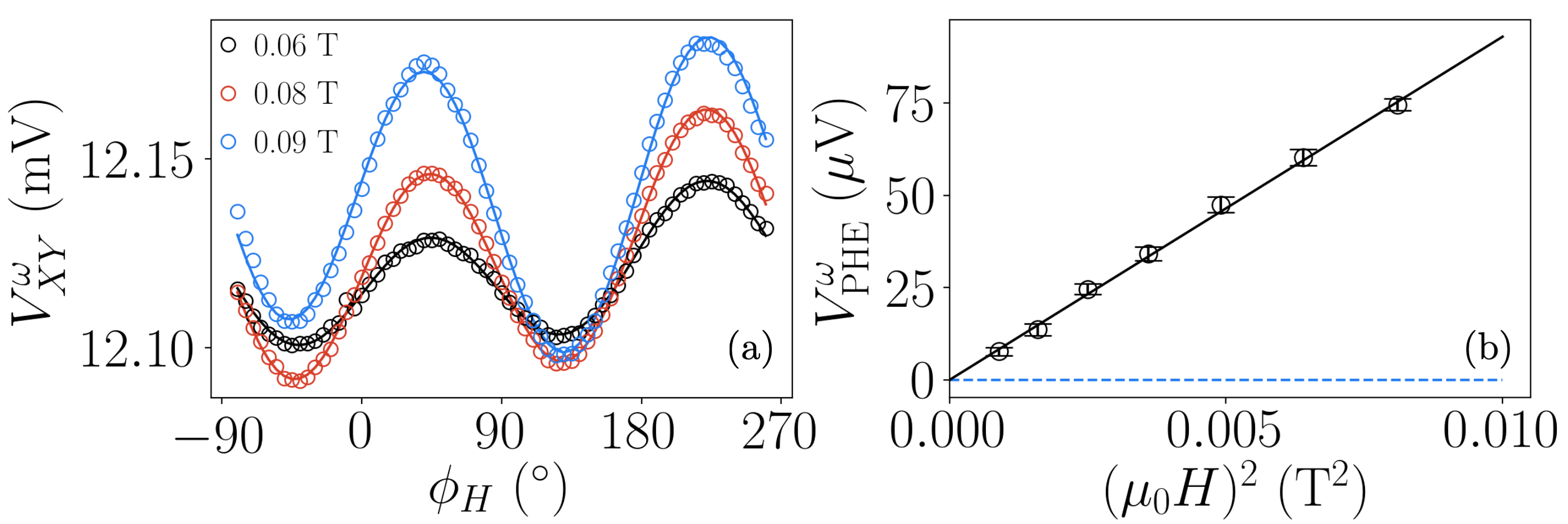}
    \caption{{\bf (a)} First-harmonic PHE data measured on the Pt(4 nm)/Co(1.15 nm) device ($\mu_0 \meff \approx 0.42$ T). The lines overlayed are best fits to equation (\ref{PMAAng}). {\bf (b)} The amplitude of the PHE signal in (a) vs. $(\mu_0 H)^2$. The line is a best fit that goes through the origin.}
    \label{PHECal}
\end{figure}

\begin{figure*}[t]
    \centering
    \includegraphics[width=\linewidth]{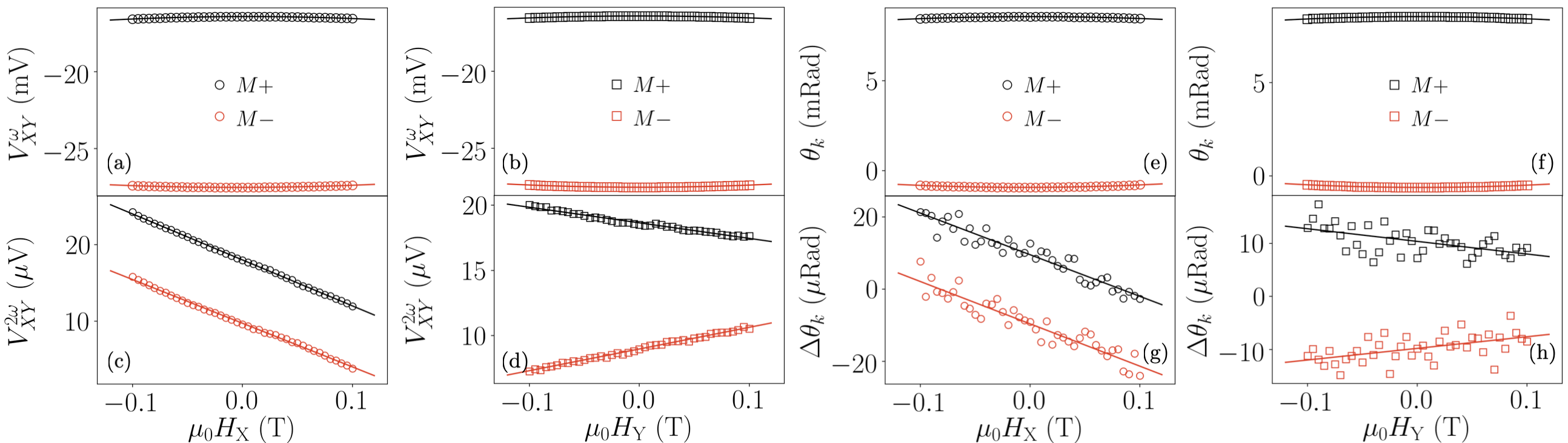}
    \caption{Measured HH and optical tilting data collected on a Pt(4 nm)/Co(1.15 nm) device with $\mu_0 \meff \approx 0.42$ T and current amplitude $\Delta I = 15$ mA. First-harmonic Hall data as a function of magnetic field swept {\bf (a)}: along the current direction and {\bf (b)}: perpendicular to the current direction. Second-harmonic Hall data as a function of magnetic field swept {\bf (c)}: along the current direction and {\bf (d)}: perpendicular to the current direction. Equilibrium Kerr rotation $\theta_k$ as a function of magnetic field swept {\bf (e)}: along the current direction and {\bf (f)}: perpendicular to the current direction. Current-induced change in the Kerr rotation  $\Delta\theta_k$ as a function of magnetic field swept {\bf (g)}: along the current direction and {\bf (h)}: perpendicular to the current direction. The second-harmonic Hall and $\Delta\theta_k$ data for the two different magnetic configurations are offset for clarity. } 
    \label{tilt}
\end{figure*}

For each sample we calibrate the anomalous Hall coefficient $R_\text{AHE}$ by measuring the change in Hall resistance upon magnetic switching as a function of out-of-plane magnetic field.  
%All the samples we report have square out-of-plane hysteresis loops indicating out-of-plane mangetic anisotropy.  
The constant of proportionality $\kappa$ relating $m_Z$ to the Kerr-rotation angle is calibrated similarly (Fig.\ 1).  To calibrate $\Rphe$, we rotate the field angle $\phi_H$ while applying a sequence of values of constant-strength in-plane magnetic field, and we measure the Hall voltage as shown in Fig.\ \ref{PHECal}(a). We determine the magnetic anisotropy term $\mu_0 M_\text{eff}$  from the first-harmonic Hall signal as a function of in-plane magnetic field swept along $\phi_H = 0$ or $\pi/2$ (see the discussion of Fig.\ \ref{tilt}(a,b) below) and then determine $R_\text{PHE}$ by fitting the measured dependence on $\phi_H$ to the form
\begin{align}
    \frac{V^{\omega}_{XY}}{\Delta I} =  &R_\text{AHE}\cos\left({\frac{H}{\meff}}\right) \nonumber\\  
    +&  R_\text{PHE}\sin^2\left({\frac{H}{\meff}}\right)\sin\phi_H\cos\phi_H \nonumber \\
   +&  R_\text{AHE}\frac{H^2\sin\theta_\text{off}}{(\meff)^2}\sin\left({\frac{H}{\meff}}\right)\cos(\phi_H - \phi_\text{off}), \label{PMAAng}
\end{align}
where the final term allows for a small misalignment of the applied field from the sample plane. The data fit well to this expected dependence -- for the sample shown in Fig.\ \ref{PHECal} with an AC current amplitude $\Delta I = 9$ mA we determine $R_\text{PHE} = 0.188(3)\ \Omega$ and $\theta_\text{off} = 0.96(2)^\circ$. Figure \ref{PHECal}(b) shows that the amplitude of the planar-Hall voltage oscillations is  proportional to $H^2$ as expected from equation (\ref{PMAAng}). The deflection angle induced over this range of applied magnetic field is in the range $\theta_0 < 15^\circ$, the same range for which the SOT measurements are performed. 

For the conversion from an effective field to spin-orbit torque efficiency (equation (\ref{taudef})), it is also necessary to calibrate the saturation magnetization $M_s$ and the current density $J_e$ in the spin-source layer. We measure $M_s$ for each heterostructure using vibrating-sample magnetometry.  We calculate $J_e$ using a parallel-conduction model after determining the thickness-dependent conductivities of the different layers in the heterostructure (See Supplementary Information section IX \cite{Supplement}).

\subsection{Electrical Detection of SOT-induced tilting}
%Our HH measurements are performed using the procedure described in refs.\ \cite{Kim2012,Hayashi2014}. We apply an AC voltage and measure the Hall voltage at the first and second harmonics as a function of an in-plane magnetic field swept in the $X$ and $Y$ directions.  The results 
The first- and second-harmonic Hall voltages measured for a Pt(4 nm)/Co(1.15 nm) device with a current amplitude $\Delta I =$ 15 mA are shown in Fig.\ \ref{tilt} for initial magnetic orientations both $m_Z = 1$ and $-1$.
%The first-harmonic Hall signal displays a parabolic dependence on  magnetic-field amplitude, with equal curvatures for the two  directions of field sweep and equal magnitudes of curvature for $m_Z = \pm1$.  The curvature reflects the  external-field-induced moment tilting away from the $Z$-axis. 
%The second-harmonic data shown in Fig.\ \ref{tilt}(c,d) are linear functions of $H_X$ and $H_Y$ as expected from equation\ (\ref{transportEq2}).  The slopes of these lines reflect the oscillatory tilting of the magnetic moment about the external-field-induced equilibrium. The lack of significant deviations from linear behavior indicates that the effective fields $\Delta H_X$ and $\Delta H_Y$ are constant to a good approximation over the range of tilt angles in the measurement.
We fit these data to equations (\ref{transportEq1}) \& (\ref{transportEq2}). From the curvature of the first harmonic we extract $\mu_0\meff = 0.424(3)$ T, which is the result used in the calibration for $R_\text{PHE}$. The second-harmonic data in Fig.\ \ref{tilt}(c,d) fit well to straight lines, indicating that the effective fields $\Delta H_X$ and $\Delta H_Y$ are constant to a good approximation over the range of tilt angles in the measurement.
From the slope of these lines and the curvature of the first harmonics, we use equation (\ref{Bdef}) to calculate that for $m_Z = -1$: $\mu_0 D_0 = -2.01(2)$ mT and $\mu_0 D_{\pi/2} = 0.62(1)$ mT, and for $m_Z = +1$: $\mu_0 D_0=2.21(2)$ mT and $\mu_0 D_{\pi/2} = 0.45(1)$ mT.  Together with the values $\Rphe = 0.188(3)\ \Omega$ and $\Rahe = 0.355(6)\ \Omega$ calibrated as described, the standard HH analysis framework (equations (\ref{TransportHX}) \& (\ref{TransportHY})) then yields the effective fields $\mu_0 \Delta H_X = 6.75(6)$ mT and $\mu_0 \Delta H_Y = -4.94(3)$ mT for the $m_Z = -1$ initial state and  $\mu_0 \Delta H_X = -6.80(4)$ mT and $\mu_0 \Delta H_Y = -4.33(3)$ mT for the $m_Z = +1$ configuration. 

\subsection{Optical Detection of SOT-Induced Tilting}

The Sagnac MOKE readouts measured simultaneously with the HH data from Fig.\ \ref{tilt}(a-d) are shown in Fig.\ \ref{tilt}(e-h). The signal-to-noise ratio for $\Delta\theta_k$ in the Sagnac measurements is not quite as high as for $V^{2\omega}_{XY}$ in the HH measurements, but it is good enough to test inconsistencies between the results of the standard HH analysis on PMA samples and the spin-orbit-torque efficiencies determined by HH measurements on in-plane samples \cite{Zhu_2019}.  A fit of the parabolic dependence of $\theta_k$ to equation (\ref{ACdeltathetak}) yields $\mu_0\meff = 0.418(3)$ T, in good agreement with value determined by HH. The values of the current-induced effective fields for this sample are determined from the slopes of the lines in Fig.\ \ref{tilt}(g,h) together with  equations (\ref{HXSagnac}) \& (\ref{HYSagnac}). For a current of $\Delta I = 15$ mA we find  
%are shown in Table \ref{tab:torques}. 
$\mu_0 \Delta H_X = 5.1(3)$ mT and $\mu_0 \Delta H_Y = -0.9(2)$ mT for the $m_Z = -1$ initial state and  $\mu_0 \Delta H_X = -5.0(3)$ mT and $\mu_0 \Delta H_Y = -0.9(2)$ mT for the $m_Z = +1$ configuration. These signs result in a positive DL SOT efficiency, $\xidl$ (consistent with literature \cite{Zhu_2019}) and a negative net FL torque, which indicates that there is a contribution from the FL torque counteracting the torque from the {\O}rsted field \cite{Ou2016}.

%Comparing to the other values displayed in this table,
%listed in Table \ref{tab:torques}, 
%We see that the Sagnac measurement is inconsistent with the standard HH analysis framework that includes the expected planar Hall signal.

\begin{figure*}[t]
    \centering
    \includegraphics[width=\linewidth]{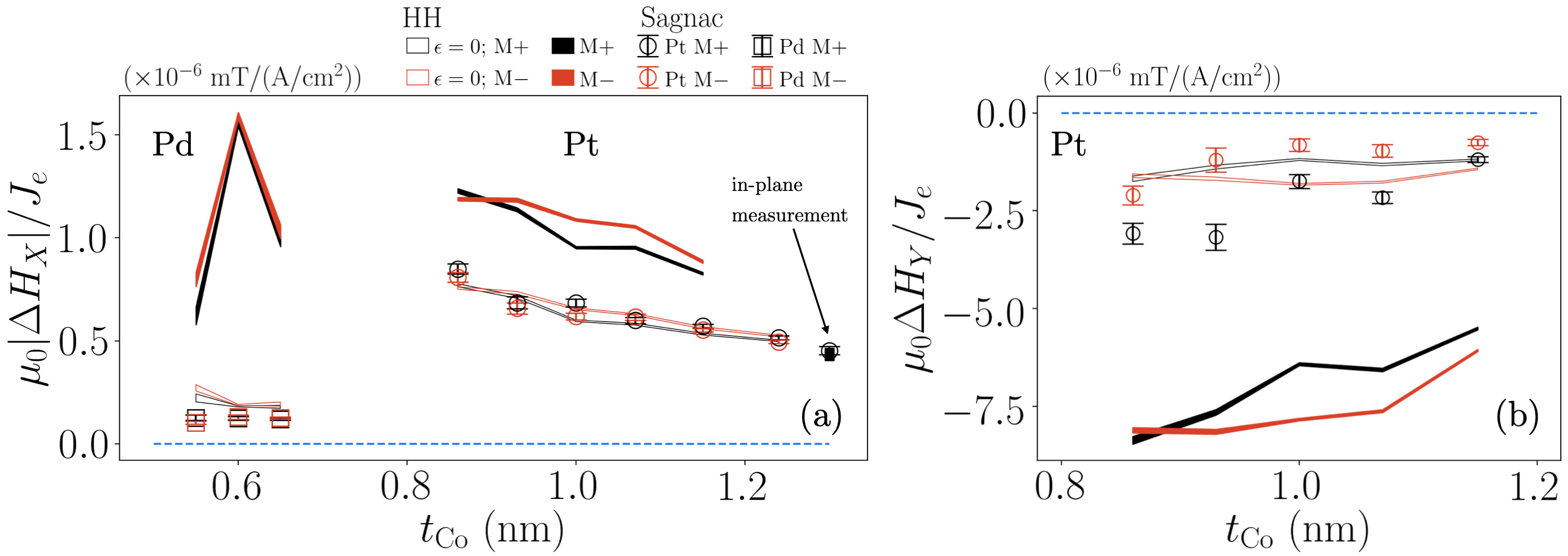}
    \caption{Calculated current-induced effective fields normalized by the current density in the Pt or Pd layer. {\bf (a)}: $\mu_0 |\Delta H_X|/J_e$ across seven devices on the Pt/Co wafer and three devices on the Pd/Co wafer. {\bf (b)}: $\mu_0 \Delta H_Y/J_e$ for devices on the Pt/Co wafer. The data points are results from the Sagnac optical measurements. Filled lines are results from the conventional HH analysis.  Empty lines are  results of a HH analysis assuming arbitrarily that $\epsilon=0$ in equations (5) and (6). The thicknesses of the lines denote 1$\sigma$ error bars.}
    \label{torques}
\end{figure*}

We have performed similar analyses for two series of Pt/Co/MgO and Pd/Co/MgO samples with different Co thicknesses. %, in order to study variations as a function of magnetic anisotropy $\mu_0 M_\text{eff}$.  
The final results for the effective fields measured by Sagnac interferometry normalized by current density flowing through the Pt or Pd are shown by the symbols in Fig.\ \ref{torques}.  To obtain these values, for each sample we measured $\Delta H_X$ and $\Delta H_Y$ for a sequence of applied voltage amplitudes and fit to a linear dependence (see equation (\ref{taudef})). (The corresponding dependences of the damping-like torque efficiency $\xidl$ on $t_\text{Co}$ are shown in Supplementary Information section VII \cite{Supplement}.)   
We compare these Sagnac results to values determined by the HH technique, for both the standard analysis that takes into account the planar Hall signal using the measured value of $\epsilon$ (filled lines) and, following the suggestion of Zhu et al.\ \cite{Zhu_2019} to arbitrarily  set $\epsilon$ = 0 in equations (\ref{TransportHX}) \& (\ref{TransportHY}) (empty lines).  The width of each line indicates the 1-$\sigma$ error bar for that sample. 
(Note in Fig.\ \ref{torques} that for the $t_\text{Co} = 1.25$ nm sample we do not present a value for the conventional HH analysis or $\mu_0 \Delta H_Y/J_e$.  Because of the relatively-weak PMA of this sample, to prevent domain formation during sweeps of in-plane magnetic field it was necessary to apply simultaneously a small constant out-of-plane magnetic field.  Our projected-field magnet was capable of performing this measurement for $\phi_H=0$ but not for $\phi_H= \pi/2$ without moving the sample.)

From Fig.\ \ref{torques} we see that for both the Pt/Co and Pd/Co samples the Sagnac results are very different from the results of the standard HH analysis that takes into account the expected planar Hall signal.  They are in much better agreement with the HH results if one assumes that the planar Hall effect somehow makes a negligible contribution to the second-harmonic Hall voltage. For the Pt/Co samples (for which $\epsilon = \Rphe/\Rahe \approx 0.5$), the standard HH analysis determines a value of $\mu_0 \Delta H_X/J_e$ that is is approximately 60\% larger than the other values, while for the Pd/Co samples (for which $\epsilon =$ 0.7 - 0.9), the standard HH framework can overestimate $\mu_0 \Delta H_X/J_e$ by as much as a factor of 15.  
% The difference in $\Delta H_Y/J_e$ for the Pt/Co samples is even greater -- $\Delta H_Y/J_e$ changes sign depending on whether or not a planar Hall contribution is assumed.  

For the Pt/Co samples, the values of the field-like component $\mu_0 \Delta H_Y/J_e$ extracted by the standard HH analysis are also in stark disagreement with the Sagnac results, while the HH analysis with $\epsilon$ arbitrarily set to 0 again agrees much better with the Sagnac values. For the Pd/Co samples, $\mu_0 \Delta H_Y/J_e$ is sufficiently weak that the uncertainties in the Sagnac measurements are comparable to the measured values, so we do not show them. 
%Overall, the Sagnac measurements demonstrate that the planar Hall effect  does not contribute to the second-harmonic Hall signal in the way that is expected within the standard analysis framework.

%To confirm the absence of a $\Rphe$ contribution to the second-harmonic Hall signals, we apply a constant in-plane field and vary the angle of the field with respect to the current direction, $\phi_H$, as shown in Fig.\ \ref{phiSweep}. In Fig.\ \ref{phiSweep}(a), we see clear evidence of a first-harmonic contribution from the PHE measured on the Pt(4)/Co(1.15) device with moderate PMA ($\mu_0\meff$ $\approx$ 0.42 T). We use the first-harmonic angular dependence to measure $R_\text{PHE}$. The second-harmonic data shown in Fig.\ \ref{phiSweep}(b), in contrast, shows no clear contribution from the PHE -- the only observable dependence is a $\cos\phi_H$ whereas the signal from the PHE rectification would contribute as a $\cos\phi_H\cos2\phi_H$ (see equation\ (\ref{transportEq2})). Fitting the second-harmonic data as shown in Fig.\ \ref{phiSweep}(c) indeed shows no measureable $V^{2\omega}_\text{PHE}$ contribution while the expectation for the $V^{2\omega}_\text{PHE}$ using the torques calculated from the field sweeps (Fig.\ \ref{torques}) and the measured $\Rphe$ (Fig.\ \ref{phiSweep}(a)) is significant. The $\phi_H$-sweeps corroborate the apparent absence of the PHE signal in the second-harmonic for the moderate-PMA device. 

\subsection{Electrical and Optical Measurements on a PMA Sample Tilted In-Plane}

Our results so far have demonstrated that the conventional HH analysis gives results inconsistent with the Sagnac measurements, but they do not prove which technique is incorrect. For that we consider additional measurements on a sample from the same wafer as our other Pt/Co/MgO devices, but with a sufficiently-thick Co layer that the PMA is weak -- specifically, we measure a Pt(4 nm)/Co(1.3 nm) sample with $\mu_0 \meff = 0.05$ T.  This weak value of PMA allows us to force the magnetization in-plane with a sufficiently-large in-plane magnetic field, and perform in-plane HH measurements as a function of the field angle $\phi_H$.  In this geometry, the current-induced damping-like effective field points out-of-plane, and it can be measured with no confusion about contributions from the planar Hall effect to first order.

\begin{figure*}[t]
    \centering
    \includegraphics[width=\linewidth]{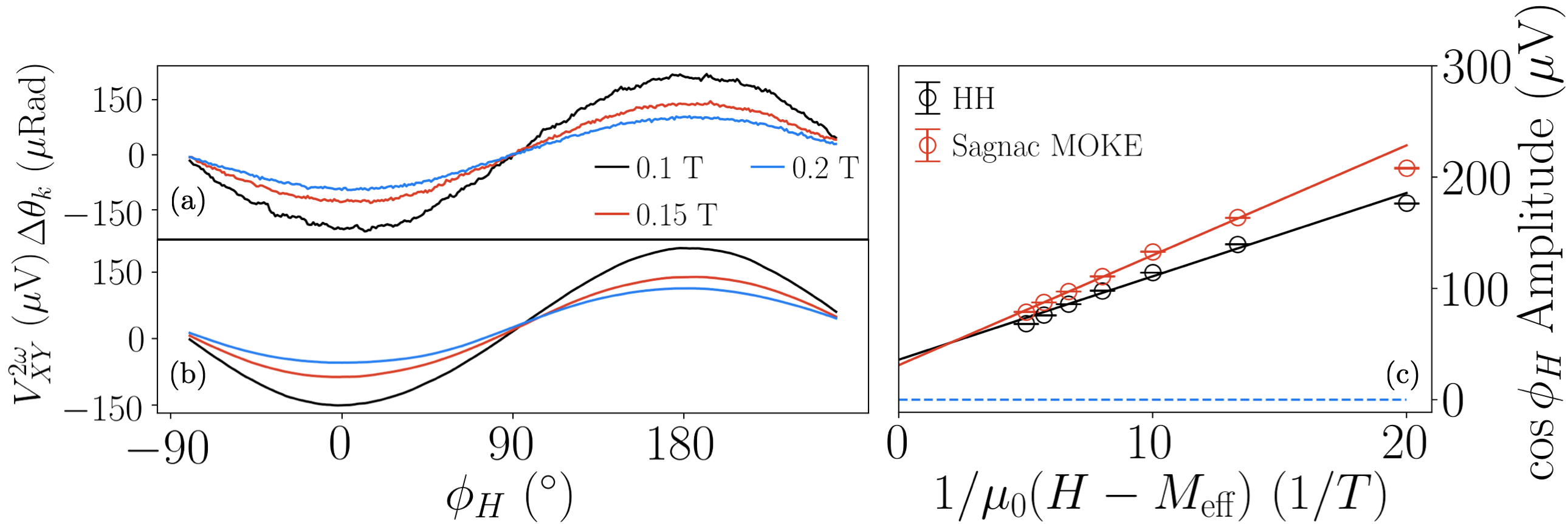}
    \caption{{\bf (a)} Second-harmonic Hall voltage ($V^{2\omega}_{XY}$) and {\bf(b)} differential Kerr rotation ($\Delta\theta_k$) measured as a function of the angle of magnetic field, $\phi_H$ for a weakly PMA Pt/Co/MgO device with $\mu_0 \meff = 0.05$ T with an applied current of $\Delta I = 13$ mA. {\bf (c)} Amplitude of the $\cos\phi_H$ components in both measurements with linear fits to to  equations (\ref{IPtiltEq}) and (14).}
    \label{weakPMAphiSweep}
\end{figure*}

Figure\ \ref{weakPMAphiSweep} shows both Sagnac MOKE and second-harmonic Hall data as a function of $\phi_H$ for this Pt(4 nm)/Co(1.3 nm) sample with a current amplitude $\Delta I=13$ mA, subject to a constant magnitude of magnetic field ($\mu_0H =$ 0.1, 0.15, and 0.2 T). We fit to the form of the signals expected for small-angle deflections in the case of an in-plane equilibrium angle \cite{Hayashi2014,Avci2014,MacNeill2017} 
\begin{align}
     \Delta\theta_k = &-\frac{\kappa\Delta H_\text{DL} \cos\phi_H}{H - \meff} \label{IPtiltEq}
\end{align}
\begin{align}
     V^{2\omega}_{XY} = &(V^{2\omega}_\text{AHE}+ V^{2\omega}_\text{ANE})\cos\phi_H + V^{2\omega}_\text{PHE}\cos\phi_H\cos2\phi_H  \\
     = &-\frac{\Delta I\Rahe\Delta H_\text{DL} \cos\phi_H}{2(H - \meff)} + V^{2\omega}_\text{ANE}\cos\phi_H \nonumber \\ 
     &- \frac{\Delta I\Rphe\Delta H_\text{FL} \cos\phi_H\cos 2\phi_H}{2H}, \nonumber \label{IPtiltXportEq}
 \end{align}
 where $V^{2\omega}_\text{ANE}$ is a voltage contribution from the anomalous Nernst effect. 
To isolate the signals due to $\Delta H_\text{DL}$, we plot the amplitude of $\cos\phi_H$ components as a function of $1/\mu_0(H-\meff)$ and perform linear fits as shown in Fig.\ \ref{weakPMAphiSweep}(c). We find $\mu_0 \Delta H_X/J_e = 4.3(3)\ (\times 10^{-14}$ T/(A/m$^2$)) from the HH measurement and $\mu_0 \Delta H_X/J_e = 4.5(2)\ (\times 10^{-14}$ T/(A/m$^2$)) from the Sagnac MOKE measurement.  These points are included in the overall summary plot in Fig.\ \ref{torques}(a).  We observe no significant $\cos\phi_H\cos2\phi_H$ component in the HH data for this sample. This could be because $\Delta H_\text{FL}$ might simply be small for this sample due to accidental cancellation between the {\O}rsted field and the field-like torque, so we do not draw any conclusions about the contribution of the planar Hall effect to the output signal for this particular sample.  For other samples with fully-in-plane anisotropy, the planar Hall effect does contribute unambiguously to give strong $\cos\phi_H\cos2\phi_H$ signals for in-plane second-harmonic Hall measurements (see, e.g., \cite{Karimeddiny2020}).   

The results of the in-plane HH and Sagnac measurements for the weakly-PMA device agree well with one another. They are also consistent with the extrapolation of the Sagnac measurements from the PMA samples to a Co thickness of 1.3 nm, but they are considerably less than expected from an extrapolation of the conventional HH analysis for the PMA samples (Fig.\ \ref{torques}(a)). Based on this we argue that the conventional HH analysis that includes the expected contribution from the planar Hall effect is incorrect. We also note that if we arbitrarily ignore the expected planar Hall contribution to the HH experiment by seting $\epsilon = 0$ in equations (\ref{TransportHX}) and (\ref{TransportHY}), then the results of the PMA HH measurements become reasonably consistent with all of the other measurement techniques. 

To be more quantitative, we compare the measured values of the damping-like torque efficiency $\xidl$ between different samples and different measurement techniques.  Unlike the current-induced effective fields, $\xidl$ is expected to be approximately independent of $t_\text{Co}$, and indeed we find this to be the case for the strong-PMA samples (see Supplementary Fig.\ 7 \cite{Supplement}). Table 1 compares the average value of $\xidl$ extracted from the HH measurements on the strong-PMA samples (using both the measured value of $\epsilon$ and then arbitrarily setting $\epsilon = 0$) to the Sagnac-MOKE measurements on the strong-PMA samples, as well as to the HH and Sagnac-MOKE measurements on the weakly-PMA sample.  Clearly, the outlier is the conventional HH analysis that includes the expected signal from the planar Hall effect.

 \begin{table}[h]
     \centering
     \begin{tabular}{c|cc}
     \toprule
       $\xidl$ & HH & Sagnac MOKE \\
       \hline
        strong-PMA tilting  & 0.23(1) & 0.146(8)\\
        strong-PMA tilting ($\epsilon=0$)  & 0.145(6) & $-$\\
        weak-PMA angle-dependence  & 0.127(7) & 0.132(6)\\
     \toprule
     \end{tabular}
     \caption{Comparison of the dampinglike spin-orbit torque efficiencies $\xidl$ measured on strong-PMA devices using small-angle tilting from an initial out-of-plane magnetic orientation (Fig.\ \ref{torques} and Supplementary Fig.\ 7 \cite{Supplement}) with values measured on a weakly-PMA sample using small-angle tilting from in-plane initial configurations (Fig.\ \ref{weakPMAphiSweep}).}
     \label{AngComparison}
 \end{table}
%The averaged $\xidl$ measured via small-angle HH tilting on the strongly-PMA devices is in stark disagreement with the $\xidl$ measured via angle-dependence on the weakly-PMA device, but reasonable agreement between HH small-angle tilting on the strongly PMA samples and the HH angle-dependence on the weakly-PMA sample is restored if we set $\epsilon=0$.
 
\subsection{Discussion}
What is wrong about the standard framework for analyzing HH measurements of PMA samples, that it yields values for the current-induced effective fields that differ from the other techniques?  Why does arbitrarily ignoring the expected planar Hall signal (i.e., arbitrarily setting $\epsilon = 0$ in equations (\ref{TransportHX}) \& (\ref{TransportHY}) in the HH analysis) give results in better agreement with these other methods?

We have considered whether the form of the current-induced effective fields might differ from the standard assumption that $\Delta H_X$ and $\Delta H_Y$ are approximately constant in the neighborhood of equilibrium tilt angles near $\theta_0 =0$.  If the current-induced effective fields were purely polar, so that there was no in-plane component to the current-induced magnetic deflections, this could explain the lack of a contribution to the second-harmonic Hall voltages from the planar Hall effect for $\phi_H = 0$ and $\pi/2$.  However, we believe that this is unphysical.  The HH results on the strongly-PMA samples imply that current-induced effective fields extrapolate to non-zero values at $\theta_0 =0$, so if they were purely polar this would require a an unphysical discontinuity. A purely-polar effective field would furthermore alter the dependence of the HH measurements on $\phi_H$ for values other than 0 and $\pi/2$, making them inconsistent with our angle-dependent measurements (Supplementary Fig.\ 5).

We have also considered whether the PMA samples might possess a nonlinear-in-current Hall effect not associated with magnetic dynamics that might largely cancel the signal expected from the PHE read-out of the current-excited magnetic dynamics.  Nonlinear-in-current Hall effects have been detected in topological-insulator-based devices \cite{Yasuda2017,He2019} and might also arise from heating-induced Nernst signals.  We suggest that this possibility deserves further analysis for heavy-metal-based structures, but we would find it a curious coincidence if a mechanism of this sort could approximately cancel the planar-Hall readout signal of spin-orbit torques in both the Pt/Co and Pd/Co devices. 

We therefore conclude that the error in the standard HH analysis is most likely in the read-out mechanism involving the planar Hall effect.  Our experiments suggest that for our PMA samples magnetic deflection induced by an applied current does not produce the same change in planar Hall resistance as the same magnetic deflection produced by an applied magnetic field.  We do not claim that the contribution of the planar Hall effect to HH signals of current-induced magnetic deflection in PMA samples is necessarily exactly zero, but it does appear to be far smaller than expected based on calibration of the planar Hall effect using magnetic-field-induced magnetic deflection -- and negligible to a good approximation.

We do not yet have a good microscopic explanation for why the planar Hall effect should not contribute to second-harmonic Hall signals for PMA samples while it does for samples with in-plane anisotropy \cite{Karimeddiny2020}.  We can speculate that magnetic tilting associated with spin-orbit torques will involve non-equilibrium spin-accumulations that are not present for magnetic-field-induced magnetic tilting, and that perhaps such spin accumulations might affect the Hall signal.  In any case, this puzzle highlights that we still lack a basic understanding about fundamental aspects of interactions among charge currents, spin currents, and ferromagnets.

\section{Conclusion}
We report measurements of current-induced torques in PMA Pt/Co/MgO and Pd/Co/MgO samples performed by simultaneously detecting small-angle current-induced magnetization tilting using both harmonic Hall (HH) measurements and Sagnac MOKE interferometry.  We find that the conventional HH analysis, which takes into account the expected read-out signals due to the planar Hall effect, is inconsistent with the Sagnac MOKE results.  The Sagnac measurements for the damping-like torque in the PMA samples are, however, consistent with both harmonic Hall and Sagnac measurements on a weakly-PMA sample forced to an initial in-plane orientation by an applied magnetic field.  These results indicate that the conventional harmonic Hall analysis for PMA samples, used in hundreds of published papers, gives incorrect values for spin-orbit torques in samples for which the planar Hall effect is significant. (For materials in which the magnetic-field-induced planar Hall effect is negligible, we do not claim any problem.) We find phenomenologically that the conventional HH analysis for PMA samples can be improved, yielding results in better agreement with other measurement techniques, by arbitrarily ignoring the expected signal from the planar Hall effect (i.e., arbitrarily setting $\epsilon$ equal to zero in equations (\ref{TransportHX}) \& (\ref{TransportHY})). Our findings help to explain previous reports of apparently-unphysical results from the conventional HH analysis \cite{Woo2014,Torrejon2014,Lee2014,Lau_2017,Zhu2019,Zhu_2019}.  We do not yet have a microscopic understanding of why current-induced magnetization tilting produces a negligible planar Hall signal in PMA samples, while the same magnetization tilting produced by an applied magnetic field does generate a planar Hall effect.
%{\color{red}[Not edited yet.]} We have performed HH measurements of SOTs on PMA magnetic heterostructures while simultaneously measuring magnetic deflection optically using MOKE that is read out to a high sensitivity by a modified Sagnac interferometer. Modeling with a conventional understanding of the PHE suggests that the PHE contributes to the second-harmonic Hall voltage; however, the corroborating MOKE measurements agree with the harmonic Hall only when the PHE contributions are diminished significantly or ignored altogether.  Additionally, when we sweep the angle of applied magnetic field, $\phi_H$, we see no sign of the angular dependence that would result from PHE-rectified signal. Strong DL and FL torques are measured when they couple to the AHE or MOKE, suggesting that it is indeed the PHE itself that is diminished and not the torques which may couple to it.

%\newpage

\section{Materials and Methods}
\subsection{Sample fabrication}
The sample heterostructures are grown by DC-magnetron sputtering at a base pressure of less than 3$\times10^{-8}$ torr on high-resistivity, surface-passivated Si/SiO$_2$ substrates. Hall bars are patterned using photolithography and ion mill etching, then Ti/Pt contacts are deposited using photolithography, sputter deposition, and liftoff. The Co is deposited with a continuous thickness gradient (``wedge") across the 4-inch wafers and all devices measured have their current flow direction oriented along the thickness gradient.  The Hall-bar devices measured are 20 $\mu$m $\times$ 80 $\mu$m in size and the change in Co thickness is negligible on this scale i.e. the gradient over 80 $\mu$m is orders of magnitude smaller than the RMS film roughness. The Ta underlayer is used to seed a smooth growth of subsequent films and the MgO/Ta forms a cap to minimize oxidation of the Co layer. 

\subsection{Sagnac Interferometer Design}
Our Sagnac interferometer, modeled after those in refs.\ \cite{Xia2006,Fried2014}, is shown Fig.\ 1.
The beamline begins with a 770 nm superluminescent diode (SLED). The beam goes through a pair of Faraday isolators that provide $>65$ dB of backward isolation and prevent back-reflections into the diode that would cause intensity fluctuations and other source instabilities. Next, the beam goes through a beam splitter, polarizer, and half-wave plate (HWP) that prepare the beam polarization to be 45$^\circ$ with respect to the slow axis of a polarization-maintaining (PM) fiber into which it is focused. The beam will henceforth be discussed as an equal linear combination of two separate beams of linearly-polarized light: one polarized along the slow axis and one polarized along fast axis of the PM fiber. A fiber electro-optic phase modulator (EOM) applies time-dependent phase to the beam traveling along the slow axis: $\phi_m\sin\omega t$. The beam then travels along 15 meters of PM fiber, whereupon it is collimated and focused by a long-working-distance objective through a quarter-wave plate (QWP) and onto a sample. The QWP is oriented such that one beam is converted to left-circularly-polarized light and the other is converted to right-circularly-polarized light. The beams then reflect off of a sample, exchanging the handedness of the beams and, if the sample is magnetic, imparting both the effects of circular dichroism and circular birefringence; the latter is equivalent to a Kerr rotation of linearly-polarized light and the two beams are now exchanged. Upon reflection, the two beams (now exchanged) backpropagate and the previously-unphased beam is now phased by $\phi_m\sin(\omega(t+\tau))$ where $\tau$ is the time it takes for the light to make the round trip back to the EOM. The two beams interfere to produce homodyne intensity oscillations at the EOM frequency. The backpropagating beams are then routed by the beam splitter and focused into a broadband avalanche photodetector (APD). The APD's output voltage is measured by a lock-in amplifier that references the driving frequency of the EOM, $\omega$. To simplify the interpretation of the signal, the frequency $\omega$ is tuned such that $\omega = \pi/\tau$ \cite{Xia2006} [$2\pi(3.3477 \text{ MHz})]$ for our apparatus). To maximize the Kerr rotation signal, the phase modulation depth $\phi_m$ is set by tuning the magnitude of AC voltage applied to the EOM to be $\phi_m = 0.92$ \cite{Fried2014}. With these simplifying calibrations, the Kerr rotation signal can be expressed as (see Supplementary Information section III for a full derivation \cite{Supplement})
\begin{align}\label{theta_k}
\begin{split}
    \theta_k %= \frac{1}{2}\arctan\left[\frac{J_2(2\phi_m)V_\text{APD}^{\omega}}{J_1(2\phi_m)V_\text{APD}^{2\omega}}\right] 
    \approx \frac{1}{2}\arctan\left[0.543 \frac{V_\text{APD}^{\omega}}{V_\text{APD}^{2\omega}}\right],
\end{split}
\end{align}
where $V_\text{APD}^{\omega}(V_\text{APD}^{2\omega})$ is the APD voltage measured at the first- and second-harmonic of the EOM frequency. We quantify our Kerr rotation noise to be less than 5 $\mu$rad/$\sqrt{\text{Hz}}$ using a low power density on the sample (2 $\mu$W/$\mu$m$^{2}$), comparable to the noise in ref.\ \cite{Fried2014}. The low power ensures that the laser does not significantly heat the sample. More details can be found in the Supplementary Information sections II \& III \cite{Supplement}.

\newpage

\section{Acknowledgements}
We acknowledge helpful discussions with Chenhao Jin, Kin Fai Mak, Yan S.\ Li, Shengwei Jiang, Fei Xue, Vivek Amin, Paul Haney, and Mark Stiles, and technical assistance from Vishakha Gupta, Rakshit Jain, and Bozo Vareskic.
%; and stimulating distractions from the best pets: Leo, Tulip, and Truffle.
We thank the the LASSP graduate student machine shop and its manager, Nathan I.\ Ellis, for advising on custom-machined parts made by S.K.\ and Y.K.L. This work was funded by the National Science Foundation (DMR-1708499), the AFOSR/MURI project 2DMagic (FA9550-19-1-0390), and Task 2776.047 of ASCENT,
one of six centers in JUMP, a Semiconductor Research
Corporation program sponsored by DARPA.  Support from the NSF via tha Cornell Center for Materials Research assisted in the construction of the Sagnac interferometer (DMR-1719875). Y.K.L.\ is supported by a Cornell Presidential Postdoctoral Fellowship and T.M.C.\ by the Singapore Agency for Science, Technology, and Research.  The devices were fabricated using the shared facilities of the Cornell NanoScale Facility, a member of the National Nanotechnology Coordinated Infrastructure (supported by the NSF via grant NNCI-1542081) and the facilities of Cornell Center for Materials Research.

\section{Author Contributions}
S.K. and Y.K.L. devised the experiment, built the Sagnac apparatus, and performed the measurements. T.M.C. fabricated the devices. S.K. performed the data analysis. S.K., D.C.R., and Y.K.L. wrote the manuscript. All authors discussed the results and the content of the manuscript.
\section{Competing Interests}
The authors declare no competing interests.
%\printbibliography
% \bibliographystyle{naturemag}
%\bibliography{bibl}\clearpage
%merlin.mbs apsrev4-1.bst 2010-07-25 4.21a (PWD, AO, DPC) hacked
%Control: key (0)
%Control: author (8) initials jnrlst
%Control: editor formatted (1) identically to author
%Control: production of article title (-1) disabled
%Control: page (0) single
%Control: year (1) truncated
%Control: production of eprint (0) enabled
%
\clearpage



\section*{Supplementary material}

# Supplementary Information:

# Sagnac interferometry for high-sensitivity optical measurements of spin-orbit torque

Saba Karimeddiny,[1,†] Thow Min Cham,[1] Daniel C. Ralph,[1,2,*] and Yunqiu Kelly Luo[1,2,†]

[1]*Cornell University, Ithaca, NY 14850, USA*

[2]*Kavli Institute at Cornell, Ithaca, NY 14853, USA*

**CONTENTS**

———
† These authors contributed equally

* Correspondence email address: dcr14@cornell.edu

# I. DERIVATION OF THE SECOND-HARMONIC HALL SIGNAL

In this section, we will derive the transport signals in a way that differs from ref. [1] but arrives at the same answer. We assume a magnetic layer with perpendicular magnetic anisotropy (PMA). The moment is tilted slightly in-plane ($\theta_0 > 0$) by varying the strength of an in-plane external magnetic field, $H$ that is applied in the plane of the sample. During this external-field-induced tilting, an AC current is applied through the device and the first- and second-harmonic voltages are measured. An image of the device and measurement geometry are shown in Supplementary Fig. 1.

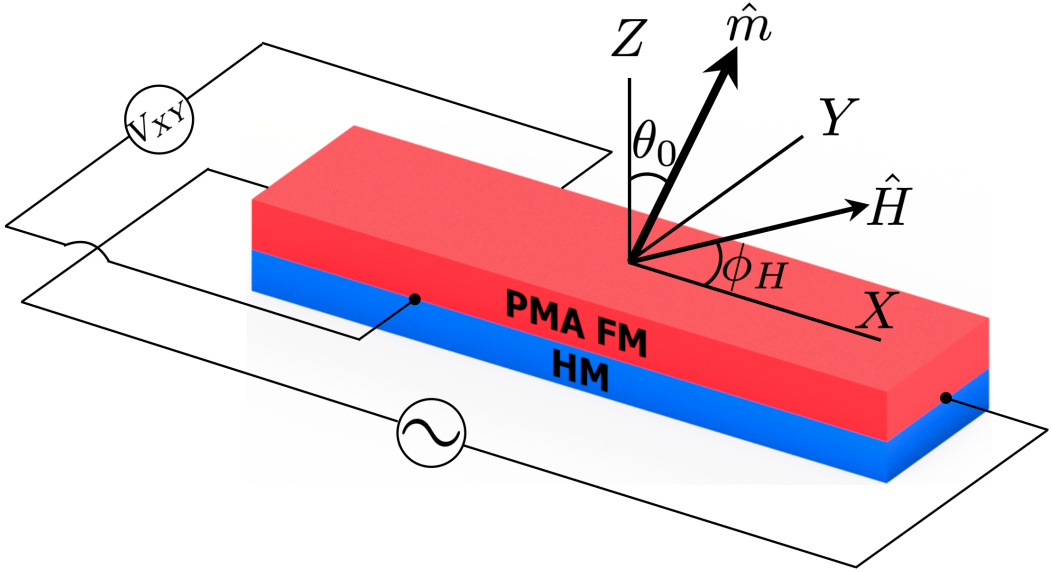

FIG. 1: The device under consideration for the PMA harmonic Hall measurements. The coordinates match those in the derivation below.

We begin the derivation by writing the equilibrium magnetic free energy divided by the total magnetic moment in the absence of any applied current

$$
\begin{aligned}
\frac{F_{\mathrm{eq}}(\theta, \phi)}{M_s} &= -\mu_0 \mathbf{m} \cdot \mathbf{H} - \frac{\mu_0 M_{\mathrm{eff}}}{2}(\mathbf{m} \cdot \hat{Z})^2 \\
&= -\mu_0 H \sin\theta \sin\theta_H \cos(\phi - \phi_H) - \frac{\mu_0}{2} \cos\theta (2H \cos\theta_H + M_{\mathrm{eff}} \cos\theta).
\end{aligned}
\tag{1}
$$

Here $F_{\mathrm{eq}}$ is the equilibrium free energy, $M_s$ is the saturation magnetization, $\mathbf{m}$ is the vector magnetic moment, $\mathbf{H}$ is the vector external magnetic field, and $\mu_0 M_{\mathrm{eff}} = 2K_\perp/M_s - \mu_0 M_s$ is the effective magnetization. PMA is indicated by a *positive* $M_{\mathrm{eff}}$. The angles in the second line denote the direction of external applied magnetic field when subscripted with an $H$ and refer to the the direction of the magnetic moment when they lack a subscript. Minimization of this free energy yields the

3

equilibrium magnetic orientation $\theta_0, \phi_0$. As we apply an AC current the SOTs produced will act as effective fields that reorient the magnetic moment. This is a "slow" process ($\dot{\mathbf{m}} \ll \gamma|H|$) so it may be described as an effective modification of equilibrium free energy (equation 1). With the perturbation from a general, current-induced effective magnetic field, $\Delta \mathbf{H}$ (assumed small compared to $H$), the free energy becomes

$$
\begin{aligned}
\frac{F(\theta, \phi)}{M_s} =& \frac{F_{\text{eq}}(\theta, \phi)}{M_s} - \mu_0 \mathbf{m} \cdot \Delta \mathbf{H} \\
\approx& \frac{F_{\text{eq}}(\theta_0, \phi_0)}{M_s} + \frac{1}{2M_s} \frac{\partial^2 F_{\text{eq}}}{\partial \theta^2}\Big|_{\theta_0, \phi_0} (\Delta\theta)^2 + \frac{1}{2M_s} \frac{\partial^2 F_{\text{eq}}}{\partial \phi^2}\Big|_{\theta_0, \phi_0} (\Delta\phi)^2 + \frac{1}{M_s} \frac{\partial^2 F_{\text{eq}}}{\partial \theta \partial \phi}\Big|_{\theta_0, \phi_0} \Delta\theta \Delta\phi \quad (2) \\
& - \mu_0 (\sin\theta \cos\phi \Delta H_X + \sin\theta \sin\phi \Delta H_Y + \cos\theta \Delta H_Z).
\end{aligned}
$$

(The first derivatives of $F_{\text{eq}}$ are zero when evaluated at the equilibrium orientation.) We have included the cross second derivative in this expression, but when evaluated it gives zero. The new magnetic orientation in the presence of the current-induced magnetic field can then be calculated as a minimization problem

$$
\frac{\partial F}{\partial \theta} = \frac{\partial F}{\partial \phi} = 0. \quad (3)
$$

We will henceforth assume that $\theta_H = \pi/2$ to match the experimental technique (the external magnetic field is applied in the plane of the device), so that $\sin\theta_0 = H/(M_{\text{eff}})$. We will also assume negligible within-plane anisotropy so that the in-plane projection of the equilibrium magnetic moment is aligned with the external field i.e. $\phi_0 = \phi_H$. The solutions of equation (3) to first order in the current-induced field yield

$$
\Delta\theta = \frac{\Delta H_Z \sin\theta_0 - \cos\theta_0 (\Delta H_X \cos\phi_H + \Delta H_Y \sin\phi_H)}{M_{\text{eff}} \cos 2\theta_0 + H \sin\theta_0} \quad (4)
$$

$$
\Delta\phi = \frac{\Delta H_X \sin\phi_H - \Delta H_Y \cos\phi_H}{H}. \quad (5)
$$

To get from the above equations to the full expected transport signal (equations (3) and (4) in the main text) we begin with the the expression for the Hall resistance

$$
R_{XY} = R_{\text{PHE}} m_X m_Y + R_{\text{AHE}} m_Z \quad (6)
$$

$$
= \frac{R_{\text{PHE}}}{2} \sin^2\theta \sin 2\phi + R_{\text{AHE}} \cos\theta. \quad (7)
$$

(Note that our definition of $R_{\text{AHE}}$ is equal to $\Delta R_A/2$ as defined in ref. [1].) We let $\theta$ and $\phi$ each consist of an equilibrium contribution due to the external magnetic field and a time-dependent contribution due to the AC-current-induced spin-orbit fields: $\theta \to \theta_0 + \Delta\theta$ and $\phi \to \phi_0 + \Delta\phi$ with

$\Delta\theta, \Delta\phi \ll 1$ and then Taylor expand

$$R_{XY} \approx \frac{R_{\mathrm{PHE}}}{2} \left( \sin^2\theta_0 + \Delta\theta\sin 2\theta_0 \right) \left( \sin 2\phi_0 + 2\Delta\phi\cos 2\phi_0 \right) + R_{\mathrm{AHE}} \left( \cos\theta_0 - \Delta\theta\sin\theta_0 \right). \quad (8)$$

We disregard the second-order terms (e.g. $\propto \Delta\theta\Delta\phi$)

$$R_{XY} \approx R_{\mathrm{PHE}} \left( \frac{1}{2}\sin^2\theta_0\sin 2\phi_0 + \Delta\phi\sin^2\theta_0\cos 2\phi_0 + \frac{1}{2}\Delta\theta\sin 2\theta_0\sin 2\phi_0 \right) + R_{\mathrm{AHE}} \left( \cos\theta_0 - \Delta\theta\sin\theta_0 \right).$$
$$(9)$$

Here we use that $\sin\theta_0 = H/M_{\mathrm{eff}}$ and $\phi_0 = \phi_H$, we substitute in the expressions for $\Delta\theta$ and $\Delta\phi$ that were derived previously, and we separate the first and second harmonics. For weak applied fields $H \ll M_{\mathrm{eff}}$ we get

$$V_{XY}^{\omega} = \left[ \pm R_{\mathrm{AHE}} \left( 1 - \frac{H^2}{2M_{\mathrm{eff}}^2} \right) + R_{\mathrm{PHE}}\frac{H^2}{2M_{\mathrm{eff}}^2}\sin 2\phi_H \right] \Delta I \quad (10)$$

$$V_{XY}^{2\omega} = \frac{1}{2}\frac{H}{M_{\mathrm{eff}}^2}\Delta I \left[ \left( \pm R_{\mathrm{AHE}} - R_{\mathrm{PHE}}\sin 2\phi_H \right) \left( \Delta H_X\cos\phi_H + \Delta H_Y\sin\phi_H \right) \right. \quad (11)$$

$$\left. + R_{\mathrm{PHE}}\cos 2\phi_H \left( \Delta H_X\sin\phi_H - \Delta H_Y\cos\phi_H \right) \right]. \quad (12)$$

The latter expression can be simplified to

$$V_{XY}^{2\omega} = \frac{1}{2}\frac{H}{M_{\mathrm{eff}}^2}\Delta I \left[ \pm R_{\mathrm{AHE}} \left( \Delta H_X\cos\phi_H + \Delta H_Y\sin\phi_H \right) \right.$$
$$\left. - R_{\mathrm{PHE}} \left( \Delta H_X\sin\phi_H + \Delta H_Y\cos\phi_H \right) \right]. \quad (13)$$

These previous equations are valid for arbitrary angles $\phi_H$ for the orientation of the in-plane applied magnetic field. When evaluated for $\phi_H = 0$ and $\pi/2$, they are equivalent to the final expressions in ref. [1] taking into account the difference in notation: $R_{\mathrm{AHE}} = \Delta R_A/2$.

## II. DETAILS OF THE SAGNAC INTERFEROMETER

We begin this section by recommending the work by Fried et al., Rev. Sci. Instrum. **85**, 103707 (2014) [2]. This paper served as the most helpful resource when building and debugging our interferometer, and many of the details in this section are inspired by the helpful level of detail in that work. We will again show our main text Fig. 1 but discuss some of the finer details of the apparatus. The entire setup, including all of the optics, sample stage, and magnet are housed on a floating optic table and enclosed in a rigid polycarbonate box affixed with sound-proof foam to block air currents and external vibrations.

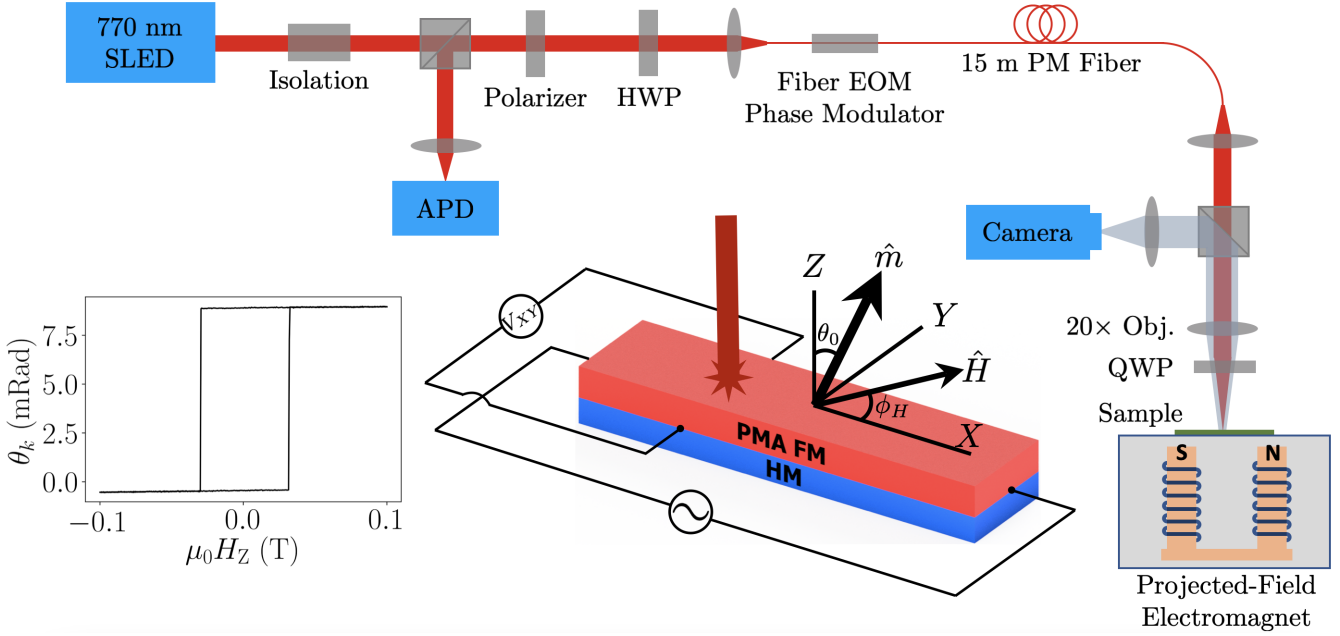

FIG. 2: Main text Fig. 1 repeated here for ease of viewing.

For the source, we use a 770 nm SLED, which has a broad ($\approx$ 15 nm) linewidth. In our original design we used a ultra-narrow-linewidth 780 nm diode, but we found that the broad-linewidth source reduced our noise by about a factor of two; this is because the small-linewidth source has a long beam coherence length. Therefore, the forward-going beam remained coherent with the reflected beam upon cycling through the apparatus, which gave them the opportunity to interfere and produce spurious interference signals not related to the Kerr rotation. The diode and most of its pigtailed fiber are stored inside a closed styrofoam box within the polycarbonate box to further prevent temperature fluctuations and air currents.

We use two Faraday isolators that provide > 65 dB of isolation to protect the diode from backreflections. The EOSpace fiber electro-optic phase modulator (EOM) is driven by a 50-MHz-bandwidth Zurich Instruments HF2LI lock-in amplifier and all of the signals (transport and optics) are detected on the same lock-in using its multiple demodulators. Our EOM is permanently pigtailed with a 5 m fiber and we append a 10 m fiber to it for a total length of 15 m. Both the EOM and the fiber are stored inside a closed styrofoam box (not the same box as the diode) within the polycarbonate box.

Upon exiting the fiber, the beam is collimated by a screw-on FC/APC lens adapter to a beam diameter of about 8 mm. We choose such a large beam diameter to maximize the filling of the back aperture of the objective lens and reduce our beam spot size on the sample. The beamsplitter after the collimating lens is retractable and it is illuminated with white light and inserted only to

align the desired sample properly under the beam. The beam does not go through this beamsplitter during measurement. For the objective we choose to use a $20\times$ near-IR ultra-long-working-distance objective lens to minimize the spot size, maximize the numerical aperture and field-of-view, and leave enough room for probes to make contact to the sample. The quarter-wave plate (QWP) is placed *after* the lens so that light is still linearly polarized while going through the lens. Most lenses have non-negligible Verdet constants so this is very important for reducing the spurious Faraday rotation incurred by the beam while it traverses the lens.

Our beam spot size on the sample is approximately 6 $\mu$m and our power incident on the device is $< 70$ $\mu$W. We find that optical powers exceeding a few hundred $\mu$W can begin to show local heating effects on the sample as indicated, e.g., as a change in the magnetic coercivity. To accommodate such a low-power beam, we detect the signal with a 50-MHz-bandwidth avalanche photodiode (APD) because it maintains a very low noise equivalent power (NEP) while sacrificing its saturation power, which we remain safely below.

## III. DERIVATION OF THE SAGNAC MOKE SIGNAL

### A. Measurement of the Kerr rotation angle $\theta_k$ in the absence of applied current

We will use the language of Jones matrices to derive equation (15) in the main text. First, we define some general Jones matrices:

$$\text{P}(\theta) = \begin{pmatrix} \cos^2\theta & \sin\theta\cos\theta \\ \sin\theta\cos\theta & \sin^2\theta \end{pmatrix} \tag{14}$$

$$\text{WP}(\theta,\phi) = \begin{pmatrix} \cos\frac{\phi}{2} + i\sin\frac{\phi}{2}\cos 2\theta & i\sin\frac{\phi}{2}\sin 2\theta \\ i\sin\frac{\phi}{2}\sin 2\theta & \cos\frac{\phi}{2} - i\sin\frac{\phi}{2}\cos 2\theta \end{pmatrix} \tag{15}$$

$$\text{EOM}(t) = \begin{pmatrix} e^{i\phi_\parallel \sin\omega t} & 0 \\ 0 & e^{i\phi_\perp \sin\omega t} \end{pmatrix} \tag{16}$$

$$\text{S} = \frac{1}{2}\begin{pmatrix} \frac{e^{-i\delta_+}}{r_+} + \frac{e^{-i\delta_-}}{r_-} & i\left(\frac{e^{-i\delta_+}}{r_+} - \frac{e^{-i\delta_-}}{r_-}\right) \\ -i\left(\frac{e^{-i\delta_+}}{r_+} - \frac{e^{-i\delta_-}}{r_-}\right) & \frac{e^{-i\delta_+}}{r_+} + \frac{e^{-i\delta_-}}{r_-} \end{pmatrix}. \tag{17}$$

Here, our Jones vectors are in the basis of the laboratory: $\text{P}(\theta)$ is a polarizer oriented at an angle $\theta$. $\text{WP}(\theta,\phi)$ is a $\phi$-wave plate oriented at an angle $\theta$. EOM is the electro-optical phase modulator that applies a voltage-dependent phase ($\phi_\perp$ or $\phi_\parallel$ depending on whether the polarization of the incoming beam is along or perpendicular-to the optical axis of the EOM crystal) at a frequency of

$\omega$. In the main text we say that the EOM only applies the phase to the beam traveling along the slow axis of the fiber; this is how the EOMs are designed, but our Jones matrix is more general to account for some phase shifts in the fast-axis beam, as well. Our final result is unchanged by this. S is the effect of the sample, which quite generally, has left- and right-circularly polarized light as its eigenvectors and applies an unequal phase ($\delta_+ \neq \delta_-$; "circular birefringence") and an unequal Fresnel reflectance ($r_+ \neq r_-$; "circular dichroism") to each of the two helicities of light. The effect of the sample reflectance exchanging the handedness of circularly polarized light is not captured by S, but will rather be accomplished by a complex conjugation later.

At the start of the beam path for the interferometer, unpolarized light exits our laser and encounters a polarizer, P, oriented such the power lost through cross-polarization of the source beam is minimized (the source diode outputs partially-polarized light). We will assume without loss of generality that polarizer angle is $0°$ so the starting point for our Jones calculus is

$$ v = \begin{pmatrix} 1 \\ 0 \end{pmatrix} . $$

From our beam path we can simply apply the time-ordered Jones matrices of our optical components:

$$ \text{P}(0)\text{WP}(\pi/8, \pi)\text{EOM}(t+\tau)\text{WP}(\pi/4, \pi/2) \left[ \text{S WP}(\pi/4, \pi/2)\text{EOM}(t)\text{WP}(\pi/8, \pi)v \right]^* . \tag{18} $$

In words, we begin with linearly-polarized light that is polarized at $0°$ ($v$), the beam goes a half-wave plate that rotates the polarization of the beam to $45°$, then goes through an EOM at time $t$, then through a quarter-wave plate, reflects from the sample, the LCP and RCP beams exchange due to the reflection (this is captured by the complex conjugation), goes through the quarter-wave plate again, through the EOM at a (now later) time $t + \tau$, and finally through the polarizer. We define $\tau$ as the time it takes for the beam to travel from the EOM to the sample and back. The result of the above matrix product is

$$ \frac{ie^{-i(\delta_+ + \delta_-)} \left( r_- e^{i(\delta_- + \phi_\perp \sin \omega t + \phi_\parallel \sin[\omega(t+\tau)])} + r_+ e^{i(\delta_+ + \phi_\parallel \sin \omega t + \phi_\perp \sin[\omega(t+\tau)])} \right)}{2\, r_+ r_-} \begin{pmatrix} 1 \\ 0 \end{pmatrix} . \tag{19} $$

In our experiment, we specifically tune EOM frequency, $\omega$, such that $\tau = \pi/\omega$ [2, 3]; this results in a substantial simplification:

$$ \left( \frac{ie^{-i\delta_- + i\phi_m \sin \omega t}}{2\, r_-} + \frac{ie^{-i\delta_- - i\phi_m \sin \omega t}}{2\, r_+} \right) \begin{pmatrix} 1 \\ 0 \end{pmatrix} \tag{20} $$

where $\phi_m = \phi_\parallel - \phi_\perp$. We detect the time-averaged intensity of light so we take half of the complex square of the above to get:

$$\frac{1}{8}\frac{1}{r_-^2} + \frac{1}{8}\frac{1}{r_+^2} + \frac{1}{8}\frac{1}{r_- r_+}\left(e^{i(\delta_+ - \delta_-)}e^{2i\phi_m \sin \omega t} + e^{-i(\delta_+ - \delta_-)}e^{-2i\phi_m \sin \omega t}\right). \tag{21}$$

We then apply the Jacobi-Anger expansion to the exponentials and let $2\theta_k = \delta_+ - \delta_-$

$$\frac{1}{8}\frac{1}{r_-^2} + \frac{1}{8}\frac{1}{r_+^2} + \frac{1}{8}\frac{1}{r_- r_+}\left(e^{2i\theta_k}\sum_{n=-\infty}^{\infty} J_n(2\phi_m)e^{in\omega t} + e^{-2i\theta_k}\sum_{n=-\infty}^{\infty} J_n(2\phi_m)e^{-in\omega t}\right). \tag{22}$$

To measure the first harmonic signal, we use a lock-in amplifier to isolate the component proportional to $\sin(\omega t)$

$$\begin{aligned}
I^\omega &= \frac{1}{T}\int_T dt \left[\frac{1}{8}\frac{1}{r_-^2} + \frac{1}{8}\frac{1}{r_+^2} + \frac{1}{8}\frac{1}{r_- r_+}\left(e^{2i\theta_k}\sum_{n=-\infty}^{\infty} J_n(2\phi_m)e^{in\omega t} + e^{-2i\theta_k}\sum_{n=-\infty}^{\infty} J_n(2\phi_m)e^{-in\omega t}\right)\right]\sin \omega t \\
&= \frac{1}{2iT}\int_T dt \left[\frac{1}{8}\frac{1}{r_- r_+}\left(e^{2i\theta_k}\sum_{n=-\infty}^{\infty} J_n(2\phi_m)e^{in\omega t} + e^{-2i\theta_k}\sum_{n=-\infty}^{\infty} J_n(2\phi_m)e^{-in\omega t}\right)\right]\left[e^{i\omega t} - e^{-i\omega t}\right] \\
&= \frac{1}{2iT}\int_T dt \left[\frac{1}{8}\frac{1}{r_- r_+}\left(e^{2i\theta_k}\sum_{n=-\infty}^{\infty} J_n(2\phi_m)\left(e^{i(n+1)\omega t} - e^{i(n-1)\omega t}\right)\right.\right. \\
&\qquad\qquad \left.\left. + e^{-2i\theta_k}\sum_{n=-\infty}^{\infty} J_n(2\phi_m)\left(e^{-i(n-1)\omega t} - e^{-i(n+1)\omega t}\right)\right)\right].
\end{aligned}$$
$$\tag{23}$$

The only terms in the sums that will survive the integration are those for which the complex time-dependent exponentials are identically 1 (i.e. when $n + 1 = 0$ or $n - 1 = 0$):

$$I^\omega = \frac{1}{2iT}\int_T dt \left[\frac{1}{8}\frac{1}{r_- r_+}\left[e^{2i\theta_k}\left(J_{-1}(2\phi_m) - J_1(2\phi_m)\right) + e^{-2i\theta_k}\left(J_1(2\phi_m) - J_{-1}(2\phi_m)\right)\right]\right] \tag{24}$$

$$= \frac{1}{T}\int_T dt \frac{1}{8}\frac{1}{r_- r_+}\left[\sin 2\theta_k\left(J_{-1}(2\phi_m) - J_1(2\phi_m)\right)\right] \tag{25}$$

$$= \frac{1}{8}\frac{1}{r_- r_+}\left[\sin 2\theta_k\left(J_{-1}(2\phi_m) - J_1(2\phi_m)\right)\right] \tag{26}$$

$$= -\frac{\sin 2\theta_k J_1(2\phi_m)}{4\,r_- r_+}. \tag{27}$$

Here we have used properties of the Bessel-$J$ functions. We can compute the second harmonic (the $\cos 2\omega t$ component) using an analogous procedure

$$I^{2\omega} = -\frac{\cos 2\theta_k J_2(2\phi_m)}{4\,r_- r_+}. \tag{28}$$

Now we can calculate $\theta_k$ and also normalize out all of the dependences on the Fresnel amplitude coefficients ($r_+$ and $r_-$) by simply taking the ratio of the two signals:

$$\theta_k = \frac{1}{2}\arctan\left[\frac{J_2(2\phi_m)I^\omega}{J_1(2\phi_m)I^{2\omega}}\right]. \tag{29}$$

For our measurements, we maximize the first harmonic signal ($\propto \theta_k$) by tuning $\phi_m$ to maximize $J_1(2\phi_m)$; this results in $\phi_m = 0.92$ [2] and $J_2(2\phi_m)/J_1(2\phi_m) \approx 0.543$. The above equation and aforementioned constant are exactly equation (15) in the main text.

### B.  Measurement of changes in the Kerr angle $\Delta\theta_k$ due to current-induced magnetic deflections

To derive a similar result with an AC applied current, we can begin at equation (22) with an added oscillation from a time-dependent $\theta_k$ that results from current-induced tilting of the magnetic moment at the current frequency $\omega_e$:

$$\frac{1}{8\,r_-r_+}\left(e^{2i(\theta_k+\Delta\theta_k\sin\omega_e t)}\sum_{n=-\infty}^{\infty}J_n(2\phi_m)e^{in\omega t}+e^{-2i(\theta_k+\Delta\theta_k\sin\omega_e t)}\sum_{n=-\infty}^{\infty}J_n(2\phi_m)e^{-in\omega t}\right). \tag{30}$$

We can apply the Jacobi-Anger expansion again

$$\frac{1}{8\,r_-r_+}\left(e^{2i\theta_k}\sum_{n,m=-\infty}^{\infty}J_n(2\phi_m)J_m(2\Delta\theta_k)e^{i(n\omega+m\omega_e)t}+e^{-2i\theta_k}\sum_{n,m=-\infty}^{\infty}J_n(2\phi_m)J_m(2\Delta\theta_k)e^{-i(n\omega+m\omega_e)t}\right). \tag{31}$$

Now we demodulate this signal at the sideband frequency $\omega\pm\omega_e$. We will only show the $\omega+\omega_e$ derivation for sign simplicity, but the result is identical for the upper and lower sidebands:

$$
\begin{aligned}
I^{\omega+\omega_e}=&\frac{1}{T}\int_T dt\,\frac{1}{8\,r_-r_+}\left(e^{2i\theta_k}\sum_{n,m=-\infty}^{\infty}J_n(2\phi_m)J_m(2\Delta\theta_k)e^{i(n\omega+m\omega_e)t}+\right.\\
&\qquad\left. e^{-2i\theta_k}\sum_{n,m=-\infty}^{\infty}J_n(2\phi_m)J_m(2\Delta\theta_k)e^{-i(n\omega+m\omega_e)t}\right)\times\cos\left(\omega t+\omega_e t\right)\\
=&\frac{1}{2T}\int_T dt\,\frac{1}{8\,r_-r_+}\left(e^{2i\theta_k}\sum_{n,m=-\infty}^{\infty}J_n(2\phi_m)J_m(2\Delta\theta_k)e^{i(n\omega+m\omega_e)t}+\right.\\
&\qquad\left. e^{-2i\theta_k}\sum_{n,m=-\infty}^{\infty}J_n(2\phi_m)J_m(2\Delta\theta_k)e^{-i(n\omega+m\omega_e)t}\right)\times\left(e^{\omega t+\omega_e t}+e^{-\omega t-\omega_e t}\right)
\end{aligned}
\tag{32}
$$

Again, to first order, the only complex exponentials that will survive integration are the ones where the exponent is identically zero. This leaves us with:

$$
\begin{aligned}
I^{\omega+\omega_e}=&\frac{1}{16\,r_-r_+}\left[e^{2i\theta_k}\left(J_{-1}(2\phi_m)J_{-1}(2\Delta\theta_k)+J_1(2\phi_m)J_1(2\Delta\theta_k)\right)+\right.\\
&\qquad\left. e^{-2i\theta_k}\left(J_{-1}(2\phi_m)J_{-1}(2\Delta\theta_k)+J_1(2\phi_m)J_1(2\Delta\theta_k)\right)\right]\\
=&\frac{1}{4\,r_-r_+}\cos 2\theta_k J_1(2\phi_m)J_1(2\Delta\theta_k).
\end{aligned}
\tag{33}
$$

In our experiments $\Delta\theta_k$ is very small so we use that $J_1(x) \approx x/2$ for small $x$

$$I^{\omega+\omega_e} = \frac{\cos 2\theta_k J_1(2\phi_m)}{4\, r_- r_+} \Delta\theta_k. \tag{34}$$

Finally, we take the ratio of this signal with the second harmonic (at $\omega$) derived earlier to reach a simple expression for the current-induced change in the Kerr signal

$$\Delta\theta_k = \frac{J_2(2\phi_m) I^{\omega+\omega_e}}{J_1(2\phi_m) I^{2\omega}}. \tag{35}$$

All of the $\Delta\theta_k$ data presented are determined using this equation.

## IV. ABSENCE OF QUADRATIC MOKE EFFECTS

In this section we justify our claim in the main text that quadratic MOKE (qMOKE), which would couple to the in-plane components of the magnetic moment, is not evident in our Sagnac data and, therefore, negligible. qMOKE signals are second-order in the device magnetization and the expected dependence of the Kerr rotation including both the polar MOKE and qMOKE signals is [4]

$$\theta_k = \kappa m_Z + \beta m_x m_y \tag{36}$$

where $Z$ is still the film normal, $m_x$ and $m_y$ (note the lowercase subscripts) are now defined such that $x$ lies along the plane of light polarization, and $\beta$ is the qMOKE coupling parameter. Since we are shining circularly polarized light on the sample we do not have a singular plane of polarization, but by decomposing each beam into linearly-polarized components we can see that we would still expect that the qMOKE introduces a phase: $e^{\pm 2i\phi_H}$ to the LCP/RCP beam, that would produce a $\pi$-periodic dependence on the in-plane field angle analogous to the symmetry of the planar Hall effect [4]. To quantify any possible coupling of light to the in-plane magnetic-moment components, we perform a measurement analogous to the calibration of the planar Hall effect discussed in the main text – we will directly measure the change in Kerr rotation as we apply an in-plane magnetic field to tilt the PMA magnet partially in plane and then rotate the field angle $\phi_H$. Supplementary Fig. 3 shows the result of this measurement for 3 different strengths of applied magnetic field. The data for this figure were collected on the same Pt(4 nm)/Co(1.15 nm) sample with $\mu_0 M_{\text{eff}} \approx 0.42$ T discussed in the main text (i.e, Figs. 1-3 in the main text). It is clear from Supplementary Fig. 4(b) that the angle dependence of Kerr rotation is not $\pi$-periodic and is moreover quite weak – there is no analog in the Sagnac data of the strong $\pi$-periodic signal in the transport data corresponding

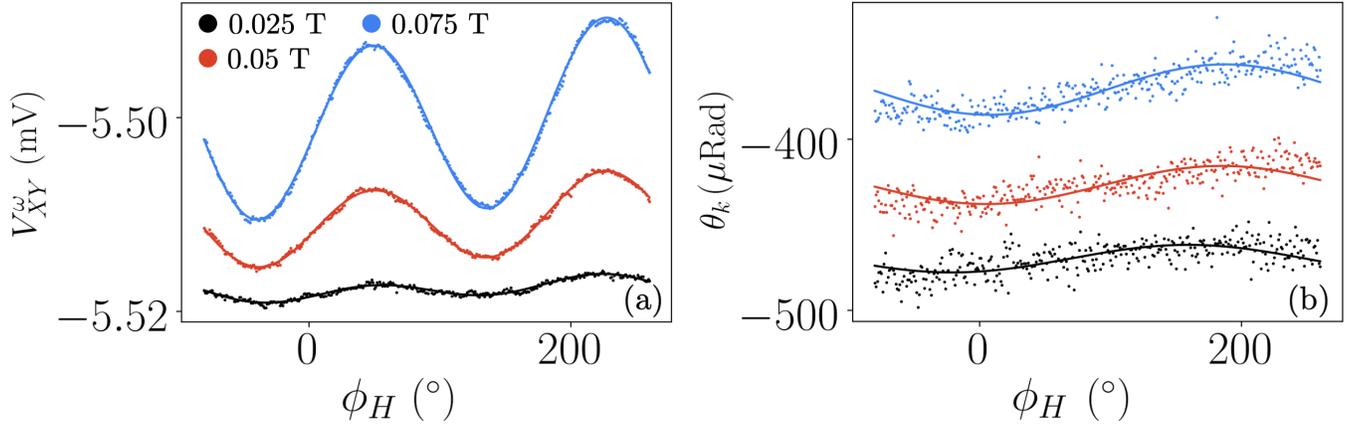

FIG. 3: **(a)** The measured first-harmonic Hall signal, $V_{XY}^\omega$ and **(b)** the measured Kerr rotation signal, $\theta_k$, vs. the angle of applied magnetic field within the plane of the magnet. Three different strengths of applied magnetic field are shown and they are artificially vertically offset in (b) for clarity. The overlayed lines are best-fits to the equations in the preceding paragraphs. The data are collected for the Pt(4 nm)/Co(1.15 nm) sample with $\mu_0 M_{\text{eff}} \approx 0.42$ T; this is the same device for which measurements are highlighted in the main text.

to the planar Hall effect. Therefore no signal corresponding to qMOKE is visible. The same holds for every other sample we have measured. We ascribe the weak $2\pi$-periodic variation in the Sagnac data to a slight misalignment of the magnetic-field relative to the sample normal. In analogy with equation (12) in the main text, for a slightly-misaligned field axis the measured Sagnac signal should have the dependence

$$
\begin{aligned}
\theta_k =& \kappa \cos\left(\frac{H}{M_{\text{eff}}}\right) \\
& + \beta \sin^2\left(\frac{H}{M_{\text{eff}}}\right) \sin\phi_H \cos\phi_H \\
& + \kappa \frac{H^2 \sin\theta_{\text{off}}}{(M_{\text{eff}})^2} \sin\left(\frac{H}{M_{\text{eff}}}\right) \cos(\phi_H - \phi_{\text{off}}),
\end{aligned}
\tag{37}
$$

Fits for both the Hall and Sagnac measurements are shown in Supplementary Fig. 3. Both sets of data indicate a field/sample tilt of $\theta_{\text{off}} \sim 1°$ (Supplementary Fig. 4). The stray field is most likely caused by a slight misalignment of the projected-field magnet (GMW 5201) center.

Fits to the Sagnac data allow us to determine an upper bound on the quadratic MOKE coefficient, $\beta \leq 3.0(3) \times 10^{-4}$ (compared to $\kappa \approx 5 \times 10^{-3}$). This limit is sufficiently small that the qMOKE cannot affect our analysis in the main text. The field-axis misalignment is also sufficiently small that it does not affect any of our quantitative conclusions.

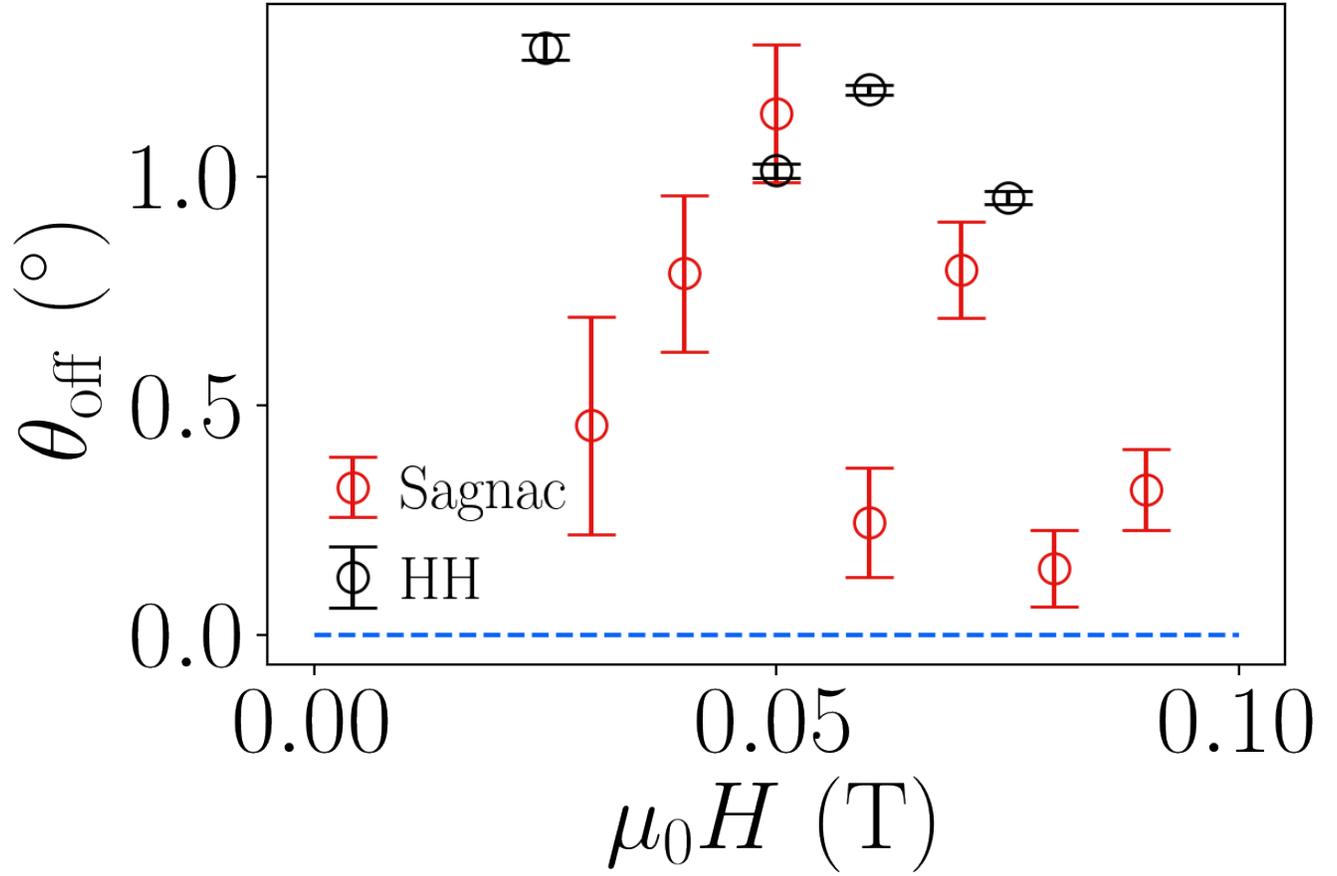

FIG. 4: The measured field misalignment angle, $\theta_{\text{off}}$, versus the strength of applied magnetic field applied nominally in the plane of the device. $\theta_{\text{off}} = 0$ means that the magnetic field is totally in the device plane.

# V. HARMONIC HALL ANGLE DEPENDENCE FOR A STRONGLY-PMA DEVICE

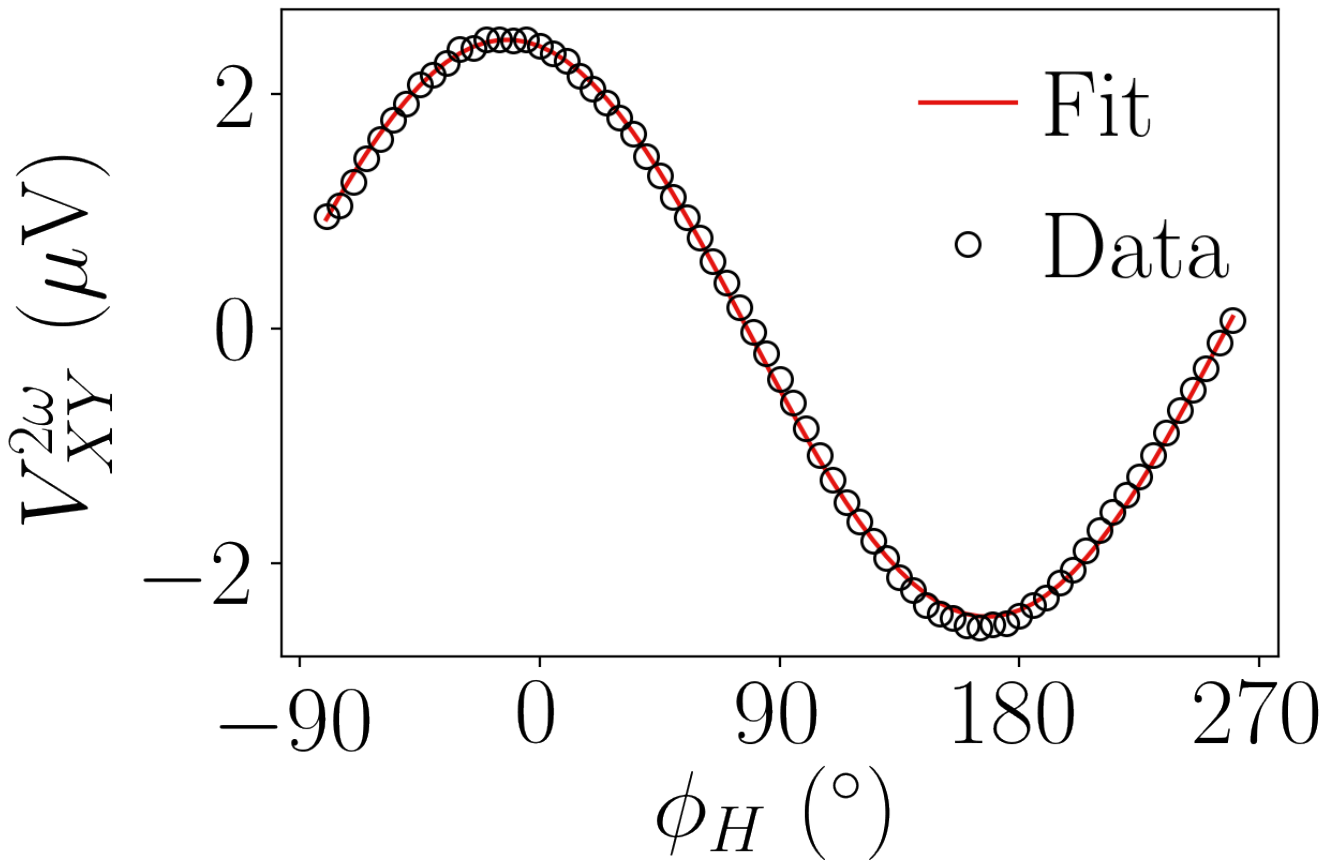

FIG. 5: The measured second harmonic Hall signal as a function of the angle of applied in-plane magnetic field. This measurement was performed on the same device that was highlighted in the main text (with $\mu_0 M_{\text{eff}} = 0.42$ T). The line is a best-fit to equation (4) in the main text. Here, $\Delta I = 9$ mA and $\mu_0 H = 0.09$ T.

# VI. VOLTAGE SWEEPS

We mention in the main text that, in order to improve our statistical uncertainty, we perform the torque measurements for a variety of applied currents in the PMA devices. We show the results of this in Supplementary Fig. 6 for the sample that was the focus of the main text figures: the Pt(4 nm)/Co(1.15 nm) sample with $\mu_0 M_{\text{eff}} \approx 0.42$ T. We see that, as expected, both the HH- and the Sagnac-measured effective fields are directly proportional to the applied current. This linearity persists for the optics measurement up to the highest applied currents, but starts to break down for high currents in the transport measurements. We will not speculate on the reason for this in

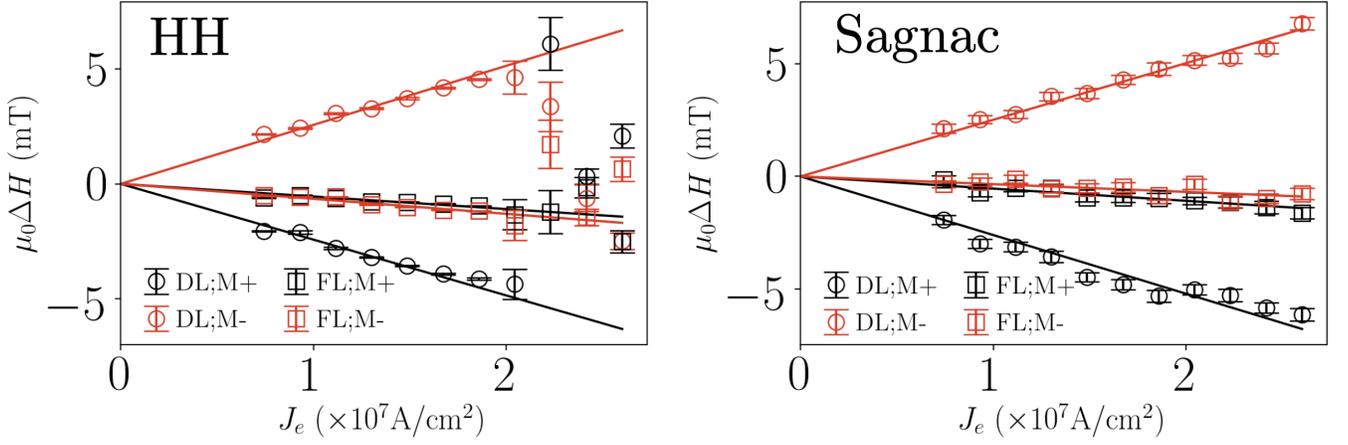

FIG. 6: The measured current-induced spin-orbit fields using both harmonic Hall transport (top) and the optical technique (bottom) as a function of current density within the Pt. The data are collected on the Pt(4)/Co(1.15) sample with $\mu_0 M_{\text{eff}} \approx 0.42$ T; this is the same device for which raw data are shown in the main text.

this work, but only note it for completeness. For the transport data, we only consider points the linear-in-current regime when fitting the lines to avoid any high-current artifacts. The slopes of the lines are the current-normalized spin-orbit fields shown in main text Fig. 5 and are related to the values of $\xi_{\text{DL}}$ shown in Supplementary Fig. 7 via the main text equation (2).

## VII. SOT EFFICIENCY

In this section we show the damping-like torque efficiency, $\xi_{\mathrm{DL}}$, for the measured Pt/Co/MgO and Pd/Co/MgO samples (Supplementary Fig. 7). These data were calculated by converting the current-induced effective fields plotted in the main text Fig. 5 to $\xi_{\mathrm{DL}}$ via the main text equation (2).

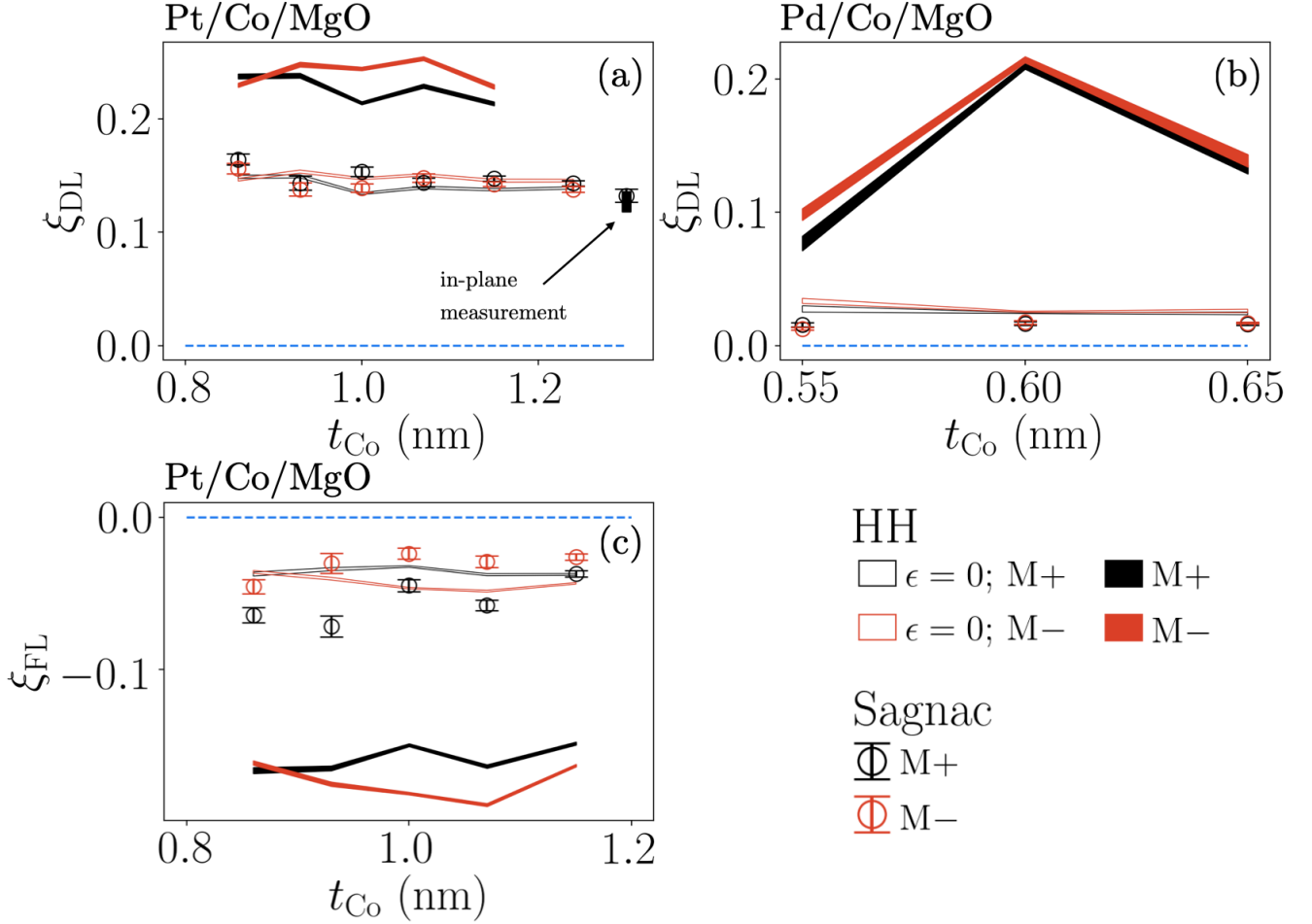

FIG. 7: **(a) & (b)** $\xi_{\mathrm{DL}}$ and **(c)** $\xi_{\mathrm{FL}}$ for the devices presented in the main text. The symbols are the results of the Sagnac interferometer measurements. The filled lines are the results of the standard analysis for the harmonic Hall measurements. The open lines are the results of the harmonic Hall measurements assuming that there is no contribution to the second-harmonic Hall signal from the planar Hall effect (i.e., that $\epsilon = 0$).

## VIII. COMPARISON OF A SAGNAC MOKE MEASUREMENT AND A CONVENTIONAL POLAR-MOKE MEASUREMENT FOR A PMA THIN FILM

Here we present a comparison between the conventional polar MOKE method and Sagnac MOKE interferometry for $Ta/Co_{40}Fe_{40}B_{20}$ test samples with perpendicular magnetic anisotropy. In Supplementary Fig. 8(a), out-of-plane magnetic hysteresis is measured using a conventional polar MOKE setup for a $Ta(4 \text{ nm})/Co_{40}Fe_{40}B_{20}(0.65 \text{ nm})$ sample. In comparison, Supplementary Fig. 8(b) shows Sagnac MOKE interferometry readout on a similar CoFeB sample from the same wedge wafer with a slightly different thickness $Ta(4 \text{ nm})/Co_{40}Fe_{40}B_{20}(0.85 \text{ nm})$. The difference in the coercivity field between Supplementary Fig. 8(a) and 8(b) is due to this small difference in film thickness. One can visually observe that the signal-to-noise of the Sagnac MOKE interferometry is a significantly improvement compared to conventional polar MOKE. The linear background in polar MOKE readout (Supplementary Fig. 8(a)) is due to the Faraday effect from the last focusing lens near the sample, while such effect is eliminated in the Sagnac MOKE interferometry by placing a quarter-waveplate between sample and the objective [2]. In addition, the Sagnac data here are a single scan taken in $\sim 1$ minute with a lock-in amplifier time constant of 10 milliseconds, while the polar-MOKE data shown here are averaged over 10 scans with a lock-in amplifier time constant of 500 milliseconds, which takes total of about 10 - 20 minutes.

Our conventional polar-MOKE setup mimics the setup of [5] with the only difference of using a Helium-neon laser as the light source. A linearly-polarized beam is incident normal to the sample through an objective lens with a numerical aperture of 0.4, focusing the beam to a circular spot with a full width at half maximum of $\sim 1 \ \mu m$. The rotation of the reflected beam polarization is detected with the use of a balanced photodiode bridge with a noise equivalent power of 1.1 pW/$\sqrt{\text{Hz}}$, which gives a readout noise of $\sim 0.4$ mRad/$\sqrt{\text{Hz}}$ for the conventional MOKE setup (Supplementary Fig. 8(a)). In contrast, our Sagnac MOKE readout noise (Supplementary Fig. 8(b)) is less than 5 $\mu$Rad/$\sqrt{\text{Hz}}$.

## IX. Pt SAMPLE DETAILS

### A. Wedge Thickness

All samples measured are grown by DC-magnetron sputtering onto a high-resistivity, surface-passivated $Si/SiO_2$ wafer. The stacks are $Si/SiO_2/Ta(1.5)/Pt(4)/Co(t_{Co})/MgO(1.9)/Ta(2)$ where all of the numbers in parentheses are layer thicknesses in nanometers. The bottom Ta is used as

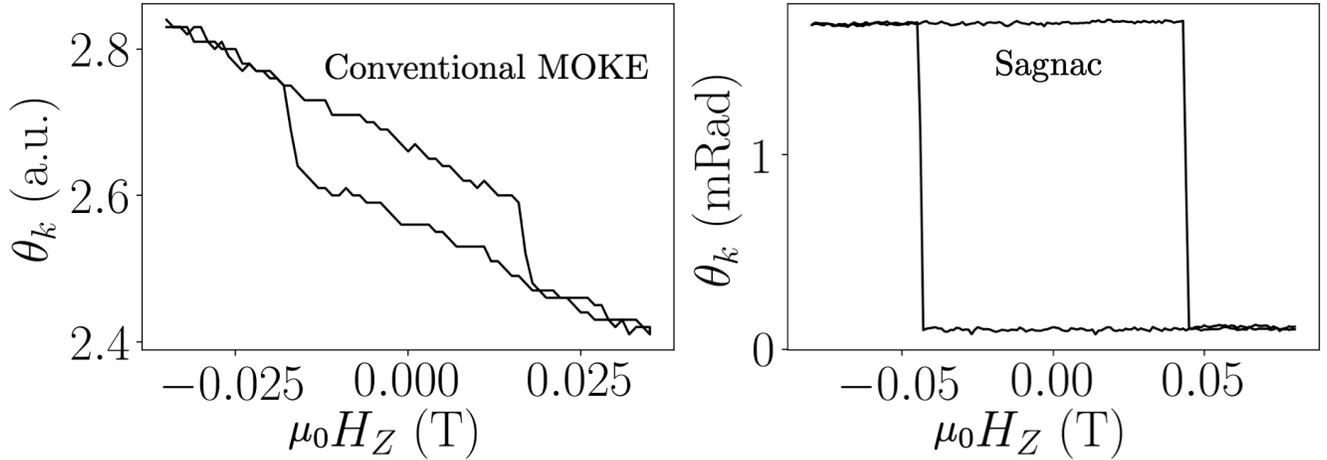

FIG. 8: **(a)** Conventional polar MOKE readout vs **(b)** Sagnac MOKE interferometry readout on CoFeB film with perpendicular magnetic anisotropy.

a seed layer to promote smooth growth of the films, and the top MgO/Ta stack is used to cap the Co and minimize oxidation of the Co Layer. Both the bottom and capping Ta layers are sufficiently resistive that they carry negligible current density compared to the Pt and Co layers. By strategically stopping the wafer rotation during sputter deposition, we grow the Co layer with a thickness-gradient "wedge". The wedge's thickness gradient is along the direction of current flow (X-axis) for all devices measured. The Co thickness as a function of device distance from the wafer flat is shown in Supplementary Fig. 9. This calibration is performed using atomic-force microscopy measurements on the thicker half of the wafer, with a subsequent polynomial fit and extrapolation to the thinnest layers.

## B. Film Conductivity

We characterize the electrical conductances of our films by measuring the four-point resistance on many devices across the Co-wedge wafer as shown in Supplementary Fig. 10. In the very low Co thickness regime ($\sim 0.4$ nm), the Co likely does not yet form a continuous film on top of the Pt so we expect the conductance measured here is entirely due to that of a bare Pt. As the Co thickness is increased, the conductance decreases due to increased surface scattering of conduction electrons in the Pt from the growing Co layer. In the regime above 0.8 nm of Co, the conductance is linear in the Co thickness, which is the expected behavior of a simple parallel-resistor model. We fit a line to the linear regime, the slope of which is the (inverse) resistivity of Co: 9.59 $\mu\Omega$cm. We estimate the resistivity of the 4 nm Pt layer adjacent to an established Co layer as corresponding approximately

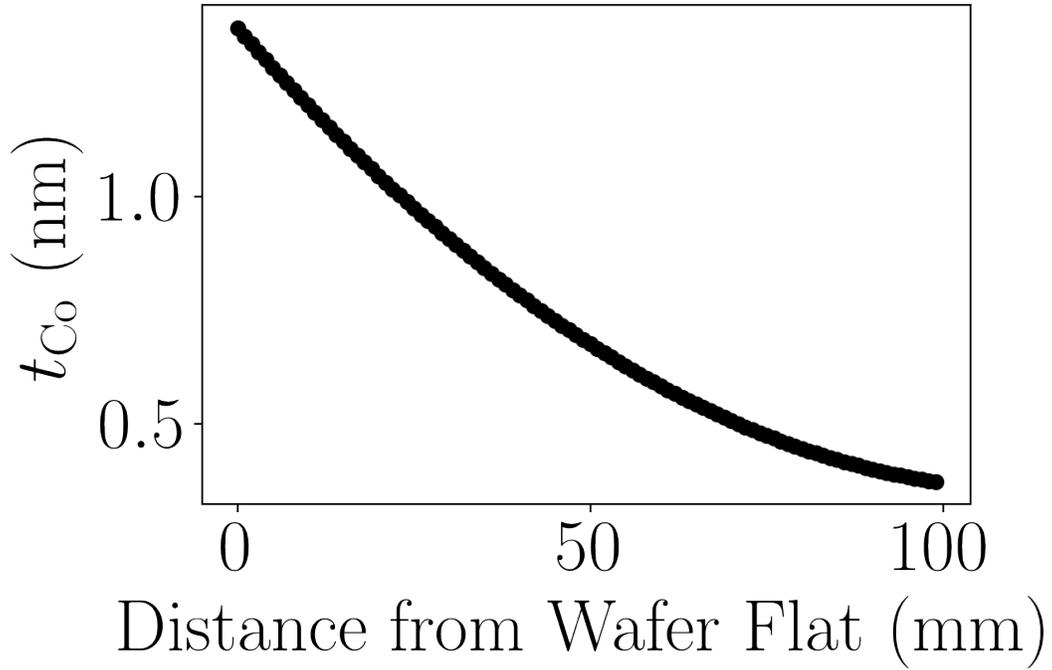

FIG. 9: The thickness of the Co "wedge" film as a function of the distance from the 4-inch wafer flat.

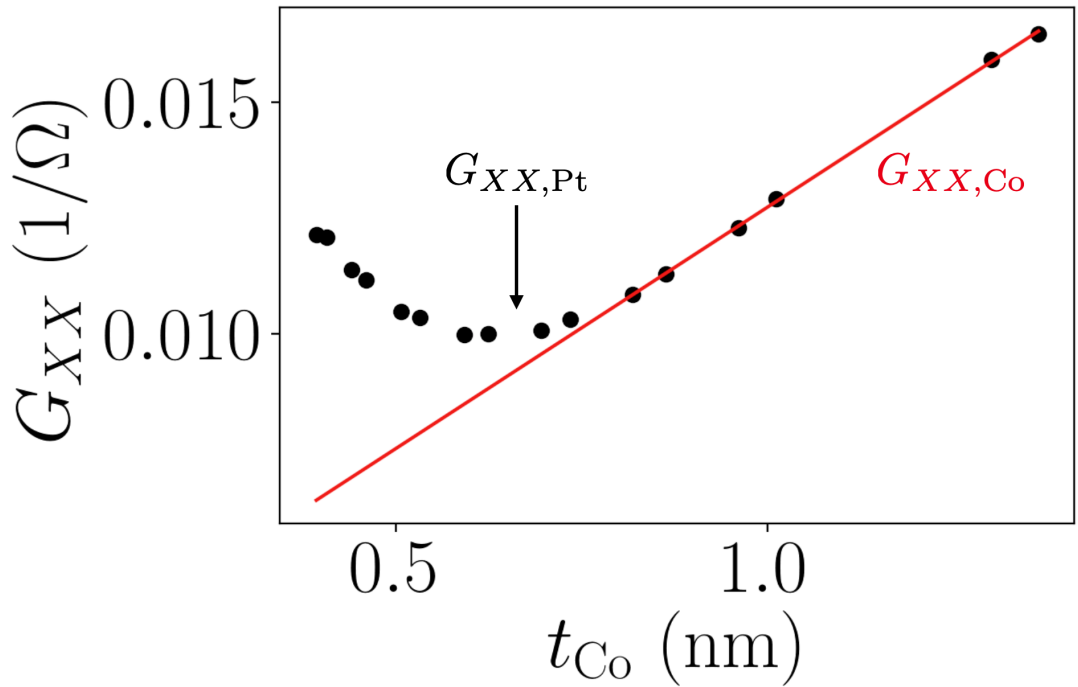

FIG. 10: Measured 4-point conductance of the fabricated devices as a function of the Co film thickness. The red line is a fit to the linear regime of the data where a parallel resistor model of the HM/FM stack is appropriate.

to the conductivity value at the minimum of the $G_{XX}$ (as indicated in the figure): 40 $\mu\Omega$cm. All of the Pt/Co/MgO samples for which we performed measurements of current-induced torque have Co layers thicker than 0.8 nm.

It is important to note that no global nor relative change in the measured resistivities of the either the Pt or the Co films could change the primary conclusions in this work: agreement of the Sagnac measurements with the $\epsilon = 0$ HH (Supplementary Fig. 7 and main text Fig. 4). Since all of the $\xi_{\text{DL}}$ are by definition normalized by the current through the Pt, any change in the resistivities would shift all of the measured $\xi_{\text{DL}}$ in concert: the pattern of agreement would be preserved.

### C. Magnetometry

We have measured the saturation magnetization of the Co in the sample stacks using vibrating sample magnetometry to be $\mu_0 M_s = 0.9$ T, and we find that the saturation magnetization is independent of the Co thickness in the thickness regime that we measured in this work.

### D. Values of $R_{\textbf{PHE}}/R_{\textbf{AHE}}$ in different Pt/Co samples

An important quantity to the HH analysis is $\epsilon = R_{\text{PHE}}/R_{\text{AHE}}$ where the total Hall magnetoresistance defined in this work is always (note the factor of 2 difference in our definition, $R_{\text{AHE}} = \Delta R_A/2$ compared to ref. [1]):

$$R_{XY} = R_{\text{AHE}} \cos\theta_0 + \frac{R_{\text{PHE}}}{2} \sin^2\theta_0 \sin 2\phi_0. \tag{38}$$

This ratio varies by less than 5% across all of the Pt/Co devices we have measured (Supplementary Fig. 11).

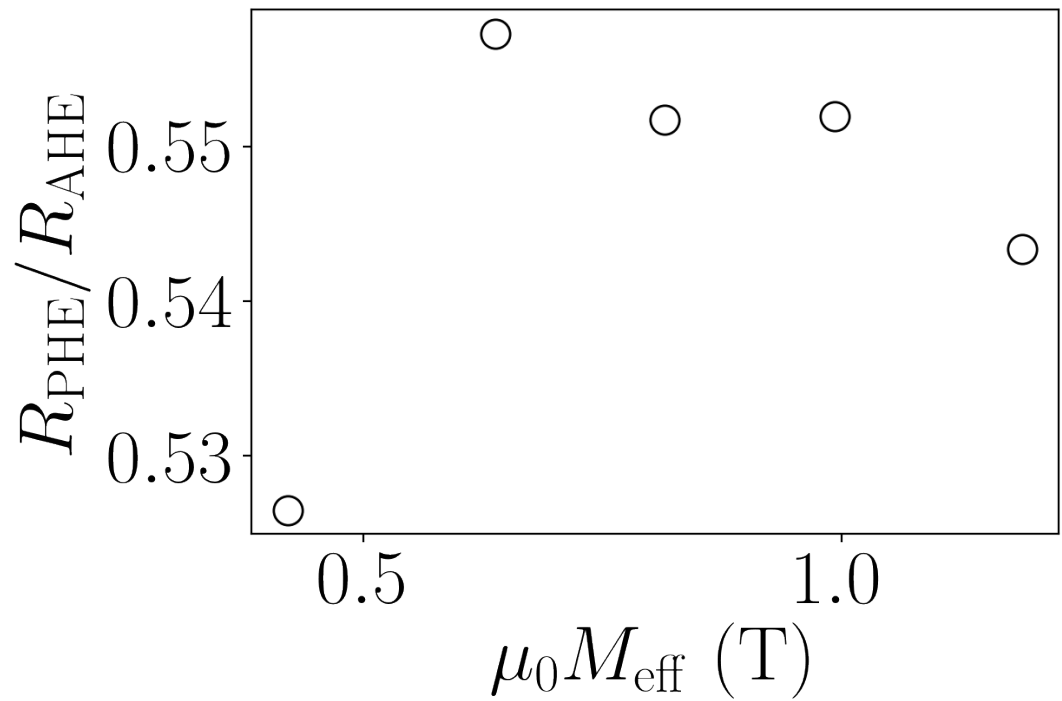

FIG. 11: The ratio $R_{\mathrm{PHE}}/R_{\mathrm{AHE}}$ as a function of $\mu_0 M_{\mathrm{eff}}$ in different Pt/Co samples.

\clearpage

\end{document}